\documentstyle[prd,aps,floats]{revtex}
\begin{document}
\draft
%
%
\input epsf
\renewcommand{\topfraction}{0.8}
\twocolumn[\hsize\textwidth\columnwidth\hsize\csname
@twocolumnfalse\endcsname
\preprint{IMPERIAL/TP/98-99/31, UAB-FT-462, hep-ph/9812533}
\title{Microwave background anisotropies in quasiopen inflation} 
\author{Juan Garc{\'\i}a-Bellido}
\address{Theoretical Physics, Blackett Laboratory, Imperial College,
Prince Consort Road, London SW7 2BZ, U.K.}
\author{Jaume Garriga and Xavier Montes} 
\address{IFAE, Edifici C,
  Universitat Aut{\`o}noma de Barcelona, E-08193 Bellaterra, Spain}
\date{December 31, 1998} 
\maketitle

\begin{abstract}
  
Quasiopenness seems to be generic to multi-field models of single-bubble
open inflation. Instead of producing infinite open universes, these
models actually produce an ensemble of very large but finite inflating
islands. In this paper we study the possible constraints from CMB
anisotropies on existing models of open inflation. The effect of
supercurvature anisotropies combined with the quasiopenness of the
inflating regions make some models incompatible with observations, and
severely reduces the parameter space of others. Supernatural open
inflation and the uncoupled two-field model seem to be ruled out due to
these constraints for values of $\Omega_0\lesssim0.98$. Others, such as
the open hybrid inflation model with suitable parameters for the slow
roll potential can be made compatible with observations.

\end{abstract}

\pacs{PACS numbers: 98.80.Cq \hspace{3mm} Preprint IMPERIAL/TP/98-99/31,
UAB-FT-462, \ hep-ph/9812533}

\vskip2pc]

\section{Introduction}

There is at present some evidence, based on observations of supernovae
at large redshift~\cite{texas1}, that the Universe may not be
Einstein-deSitter ($\Omega_{\rm m}=1,\ \Omega_\Lambda=0$), as predicted
by the simplest models of inflation. Furthermore, recent observations of
the microwave background (CMB) and large scale structure indicate that
the Universe may be open ($\Omega_0 = \Omega_{\rm m} + \Omega_\Lambda =
0.8 \pm 0.3$)\cite{texas2}. In the near future, observations of the
microwave background with a new generation of satellites, MAP and
Planck, will determine with an accuracy of order 1\% whether we live in
an open Universe or not. It is therefore crucial to know whether
inflation can be made compatible with such a Universe.

The idea that the Universe might be open is an old one, see for
instance \cite{Peebles}. Early attempts to accommodate standard
inflation in an open Universe failed to realize the fact that, in
usual inflation, homogeneity implies flatness~\cite{turner}, because
of the Grishchuck--Zel'dovich effect~\cite{GZE}. The possibility of
having, through the nucleation of a single bubble in de Sitter space,
a truly open Universe arising from inflation is not new either, see
\cite{Gott,ratra}. However, a concrete realization of a fully
consistent model was suggested only recently, the single-bubble open
inflation model~\cite{singlebubble,LM}. Since then, there has
been great progress in determining the precise primordial spectra of
perturbations~\cite{LW,sasaki,YST,Hamazaki,bubble,Garriga,Cohn,modes,TS,new,slava,quasi,gtv,gmst2},
most of it based on quantum field theory in spatially open spaces.
Simultaneously, a large effort was made in model
building~\cite{LM,Green,induced,GBL} and in constraining the existing
models from observations of the temperature power spectrum of CMB
anisotropies~\cite{induced,GBL,super,JGB}.

Open inflation models provide a natural scenario for understanding the
large-scale homogeneity and isotropy. Furthermore, these inflationary
models generically predict a nearly scale-invariant spectrum of
density and gravitational wave perturbations, which could be
responsible for the observed CMB temperature anisotropies. Future
precise observations could determine whether indeed these models are
compatible with the observed features of the CMB power spectrum. For
that purpose it is necessary to know the predicted spectrum with
great accuracy. As we will show, open models have a more complicated
primordial spectrum of perturbations, with extra discrete modes and
possibly large tensor anisotropies.

The simplest inflationary model consistent with an open Universe
arises from the nucleation of a bubble in de Sitter space~\cite{Gott},
inside which a second stage of inflation drives the spatial curvature
to {\it almost} flatness. The small deviation from flatness at the end
of inflation will be amplified by the subsequent expansion during the
radiation and matter eras. The present value of the density parameter
is determined from 
$(1-\Omega_0)/\Omega_0 \sim \exp(-2N_e)\times10^{54}$, 
where $N_e\leq 60$ is the required number
of $e$-folds during inflation (inside the bubble), in order to give an
open Universe today. A few percent change in $N_e$ could lead to an
almost flat Universe or to a wide open one.

This simple picture could be realized in the context of a single-field
scalar potential~\cite{singlebubble} or in multiple-field
potentials~\cite{LM,Green,induced,GBL}, as long as there exists a false
vacuum epoch during which the Universe becomes homogeneous and then one
of the fields tunnels to the true vacuum, creating a single isolated
bubble. The space-time inside this bubble is that of an open Universe
\cite{CDL,Gott}. Although single-field models can in principle be
constructed, they require a certain amount of fine-tuning in order to
avoid tunneling via the Hawking--Moss instanton~\cite{LM}. The problem
is that a large mass is needed for successful tunneling and a small
mass for successful slow-roll inside the bubble. For that reason, it
seems more natural to consider multiple-field models of open
inflation~\cite{LM,Green,induced,GBL}, where one field does the
tunneling, another drives slow-roll inflation inside the bubble, and
yet another may end inflation, as in the open hybrid model~\cite{GBL}.
Such models account for the large-scale homogeneity observed by
COBE~\cite{COBE} and are also consistent with recent determinations of a
small density parameter~\cite{Dekel,Wendy,Charley,CB}.

In this paper, we shall systematically explore the possible
constraints from CMB anisotropies on existing models of open and
quasiopen inflation. The structure of the paper is as follows. In
sections II and III, a brief review of the quantum tunneling and the
primordial spectrum of perturbations is given.  Section IV is devoted
to quasi\-open inflation. In section V some general bounds are found
from CMB observations. In section VI, the bounds derived in the
previous section are used to constrain several models of open
inflation. Section VII discusses the probability distribution
for $\Omega_0$ in quasi-open inflation and corresponding constraints
on the parameters of the models. Section VIII contains our
conclusions.

\section{Quantum tunneling}

The quantum tunneling that gives rise to the single bubble of open
inflation can be described with the use of the bounce action formalism
developed by Coleman--DeLuccia~\cite{CDL} and by Parke~\cite{Parke} in
the thin-wall approximation, valid when the width of the bubble wall is
much smaller than the radius of curvature of the bubble. This only
requires that the barrier between the false and the true vacuum be
sufficiently high, $U_0 \gg \Delta U=U_F-U_T$.  In this case we can
write the radius of the bubble in terms of the dimensionless parameters
$a$ and $b$~\cite{JGB},
\begin{eqnarray}\label{RHT}
R_0 H_T &=& [1+(a+b)^2]^{-1/2} \equiv [1+\Delta^2]^{-1/2}\,,\\
&& \hspace*{-5mm} a\equiv{\Delta U\over3S_1H_T},\hspace{1cm} 
b\equiv{\kappa^2S_1\over4H_T}\,,\label{ab}
\end{eqnarray}
where $\kappa^2\equiv8\pi G$ and $S_1$ is the surface tension of the
bubble wall, computed as $S_1 = \int_{\sigma_F}^{\sigma_T}
d\sigma\,[2(U(\sigma)- U_F)]^{1/2}$. Here $\sigma$ is the tunneling
field. Since $S_1\sim U_0/M \sim M(\Delta\sigma)^2$ for a mass $M$ in
the false vacuum, the parameter $a\simeq (\Delta U/U_0)\,M/H_T$, which
characterizes the degeneracy of the vacua, can be made arbitrarily small
by tuning $U_T\simeq U_F$. On the other hand, the parameter $b\simeq
(\Delta\sigma/M_{\rm Pl})^2 M/H_T$, which characterizes the width of the
barrier, is not so easily tunable, and could be very large or very small
depending on the model.

In order to prevent collisions with other nucleated bubbles (at least
in our past light cone) it is necessary that the probability of
tunneling be sufficiently suppressed. For an open Universe of
$\Omega_0 > 0.2$, this is satisfied as long as the bounce action $S_B
> 6$, see Ref.~\cite{Gott}. This imposes only a very mild constraint
on the tunneling parameters $a$ and $b$, as long as the energy density
in the true vacuum satisfies $U_T \ll M_{\rm Pl}^4$, see
Ref.~\cite{JGB}. 

\section{Primordial perturbation spectra}

There are essentially two kinds of primordial perturbations in open
inflation, subcurvature and supercurvature. In the first category we
have the usual scalar and tensor metric
perturbations~\cite{sasaki,YST,TS}, which are generated during the
slow roll inflation inside the bubble. The second category of modes
arises because of perturbations which are generated outside the bubble
or as a result of the acceleration of the expanding
bubble~\cite{LW,super,YST}. In particular, the fluctuations of the
bubble wall itself generate
perturbations~\cite{bubble,Garriga,Cohn,new} which are specific to
open inflation. In addition, in the context of two-field models of
open inflation~\cite{LM}, we have semi-classical effects due to
tunneling to different values of the inflaton
field~\cite{slava,quasi}. All these perturbations create anisotropies
in the CMB, which distort the angular power spectrum on large scales
(low multipoles). On smaller scales, the particle horizon at last
scattering subtends an angle of about one degree on the sky for a flat
universe, and somewhat smaller for an open universe, due to the
projection effect of the geodesics. This effect shifts the first
acoustic peak of the temperature power spectrum to higher multipoles
($l_{\rm peak}\sim 208\ \Omega_0^{-1/2}$)~\cite{Mark,frampton},
but the primordial spectrum at those multipoles is essentially that of
a flat Universe.

\subsection{Subcurvature scalar and tensor perturbations}

Soon after the proposal of a single-bubble open inflation
model~\cite{singlebubble}, the primordial spectrum of scalar
perturbations was computed and found to be identical to that of the flat
case, except for a prefactor that depends on the bubble
geometry~\cite{YST},
\begin{equation}\label{AS2}
{\cal P}_{\cal R}(q) = A_S^2 \,f(q)\,, \hspace{1cm} 
A_S^2 = {\kappa^2\over2\epsilon} \Big({H_T\over2\pi}\Big)^2\,.
\end{equation}
Here ${\cal P}_{\cal R}(q)$ is the primordial spectrum of scalar
metric perturbations ${\cal R}$, in term of which the continuum part
of the power spectrum\footnote{See the Appendix A for notation.
  \label{notation}} is written as
\begin{equation}\label{PRq}
\langle|{\cal R}(q)|^2\rangle = 
{2\pi^2{\cal P}_{\cal R}(q)\over q(1+q^2)}\,.
\end{equation}
The function $f(q)$ \cite{YST} depends on the tunneling parameters $a$ and
$b$, see Eq.~(\ref{ab}),
\begin{equation}\label{fq}
f(q) = \coth\pi q - {z^2\cos\tilde q + 2q z \sin\tilde q\over
(4q^2 + z^2)\,\sinh\pi q }\,,
\end{equation}
where $\tilde q = q \ln((1+x)/(1-x))$ and
\begin{eqnarray}
x &=& (1-R_0^2H_T^2)^{1/2} = \Delta\,(1+\Delta^2)^{-1/2}\,, 
\label{x}\\[1mm]
z &=& (1-R_0^2H_T^2)^{1/2}-(1-R_0^2H_F^2)^{1/2} \nonumber\\
  &=& 2b\,(1+\Delta^2)^{-1/2}\,, \label{z}
\end{eqnarray}
see Eq.~(\ref{RHT}). The function $f(q)$ is linear at small $q$, and
approaches a constant value $f(q) = 1$ at $q\geq2$. Here $q$ is the
effective momentum for scalar modes in an open Universe, determined
from $q^2=k^2-1$, where $-k^2$ is the eigenvalue of the Laplacian. For
scalar perturbations, the effect of $f(q)$ on the temperature power
spectrum is almost negligible, and therefore the tilt of the scalar
spectrum is approximately given by the same formula as in flat
space~\cite{LL93}:
\begin{equation}\label{nS}
n_S - 1 \equiv {d\ln{\cal P}_{\cal R}(k)\over d\ln k} \simeq
- 6\epsilon + 2\eta \,,
\end{equation}
in the slow-roll approximation,
\begin{equation}\label{slowroll}
\epsilon = {1\over2\kappa^2}\,\Big({V'(\phi)\over V(\phi)}\Big)^2
\ll 1 \,, \hspace{8mm}
\eta = {1\over\kappa^2}\,{V''(\phi)\over V(\phi)} \ll 1 \,.
\end{equation}

On the other hand, the primordial spectrum of tensor or gravitational
waves anisotropies took much longer to evaluate. For some years, there
seemed to be a problem with the power spectrum of tensor anisotropies,
which presented an infrared divergence at $q=0$ in an open
Universe~\cite{Allen}. It was clear that a physical regulator was
necessary. But this is precisely the role played by the bubble wall;
recent computations have shown that in the presence of the bubble the
tensor primordial spectrum\ref{notation} is given by~\cite{TS} 
\begin{equation}
\langle|h(q)|^2\rangle = \frac{\pi^2{\cal P}_g(q)}{4 q(1+q^2)}\,,
\end{equation}
where
\begin{equation}
{\cal P}_g(q) = A_T^2 \,f(q) \,, \hspace{1cm} \label{AT2}
A_T^2 = 8\kappa^2 \Big({H_T\over2\pi}\Big)^2\,.
\end{equation}
It is the shape of the function $f(q)$, see Eq.~(\ref{fq}), that gives a
finite physical observable, as we will explain in
Section~\ref{scaten}. Here $q$ is the effective momentum for tensor
modes, defined by $q^2=k^2-3$.  While for scalar modes the presence of
this function $f(q)$ in the primordial spectrum becomes irrelevant for
observations, for tensor modes the slope of the function at $q=0$ is an
important ingredient in the final value of the predicted power spectrum
at low multipoles~\cite{JGB}.

Note that the above spectra have been obtained under the small
backreaction approximation.  Immediately after nucleation, the scalar
and tensor modes have been assumed to evolve in a nearly de Sitter
spacetime. This assumption is reasonably fulfilled in the models
considered in this paper, but in models where the inflaton moves
initially very fast, the infrared end of the spectrum may be slightly
different\cite{gmst2}.

\subsection{Supercurvature and bubble-wall modes}

Apart from the usual continuum ($q^2\geq0$) of scalar and tensor modes,
generalized to an open Universe, in single-bubble inflationary models
there is a new type of quantum fluctuations, which basically account for
the combined effect of fluctuations generated during the false vacuum
dominated era which penetrate the bubble, as well as excitations of the
slow roll field generated by the accelerated growth of the bubble.
These modes are discrete modes, which appear in particular when the mass
of the scalar field in the false vacuum is smaller than the rate of
expansion, $m^2<2H^2$.  The supercurvature mode was first postulated in
Ref.~\cite{LW}, from purely mathematical arguments related to
homogeneous Gaussian random fields in spatially open universes. Only
later did concrete models of single-bubble open
inflation~\cite{sasaki,YST} show its existence in the primordial
spectrum of scalar fluctuations. These modes are characterized by having
an imaginary effective momentum (or equivalently, $k^2<1$), and
therefore describe fluctuations over scales larger than the curvature
scale, whence its name of {\it supercurvature} mode [remember than the
curvature scale corresponds to the eigenmode with eigenvalue 
$k^2=1, \ q^2=0$].

The amplitude of the supercurvature mode (for the usual two-field
models like the ones described by (\ref{general}), see below) is found
to be\cite{YST,super}~\ref{notation}
\begin{equation}
\langle| {\cal R}_{\Lambda}|^2\rangle =  \pi^2 A_{SC}^2 \,,
\end{equation}
where
\begin{equation}\label{ASC}
A^2_{SC} = {\kappa^2\over\epsilon}\,\Big({H_F\over2\pi}\Big)^2 =
A^2_S \, {2H_F^2\over H_T^2} \,,
\end{equation}
whith $A^2_S$  given by Eq.~(\ref{AS2}).

However, this is not the only discrete supercurvature mode possible in
single-bubble models. As realized in
Refs.~\cite{Hamazaki,bubble,Garriga}, there are also scalar
fluctuations of the bubble wall with $k^2=-3, \ q^2=-4$, which could
in principle have its imprint in the CMB anisotropies. Their amplitude
was computed in Refs.~\cite{Garriga,Cohn}, based on field theoretical
arguments:
\begin{equation}
\langle|{\cal R}_W|^2\rangle = 4 \pi^2 A_W^2\,.
\end{equation} 
Here
\begin{equation}\label{A2W}
A^2_W = {\kappa^2\over2z}\,\Big({H_T\over2\pi}\Big)^2
= A_S^2\,{\epsilon\over z}\,,
\end{equation}
where $z$ is given by Eq.~(\ref{z}) and $A^2_S$ is the scalar
amplitude (\ref{AS2}). Such modes contribute as transverse traceless
curvature perturbations~\cite{Garriga,bubble}, which nevertheless
behave as a homogeneous random field~\cite{modes}. It was later
realized~\cite{new} that the bubble-wall fluctuation mode is actually
part of the tensor primordial spectrum, once the gravitational
backreaction is included, so we should not consider it as a new
source of anisotropies if the tensor mode is properly taken into
account. 

\section{Quasiopen inflation}

All single-field models of open inflation predict the above primordial
spectra of anisotropies: a continuum of scalar and tensor modes and
the (possible) discrete supercurvature modes. However, as
mentioned in the introduction, it is difficult to construct such
models without a certain amount of fine-tuning~\cite{LM}, and thus
multiple-field models were considered~\cite{LM,Green,induced,GBL}.
In these models, one field $\sigma$ would do the tunneling while 
inside the bubble a second field $\phi$ would drive slow-roll
inflation.
However, a large class of two-field models do not lead to infinite
open Universes, as was previously thought, but to an ensemble of
very large but finite inflating `islands'. The most
probable tunneling trajectory corresponds to a value of the inflaton
field at the bottom of its potential; large values, necessary for the
second period of inflation inside the bubble, only arise as localized
fluctuations. These fluctuations are
provided precisely by the supercurvature modes of the slow-roll
field $\phi$, which due to their long wavelength can create large
regions of size larger than the hubble radius where the field is
coherent and thus can drive inflation.
The interior of each nucleated bubble will contain an
infinite number of such inflating regions of co-moving size of order
$\gamma^{-1}$, where $\gamma\ll 1$ is given by the supercurvature
eigenvalue $\gamma=1+q^2\equiv 1-\Lambda^2$ (this in turn depends on 
the parameters of the model, see below). We
may happen to live in one of those patches of co-moving size $d\leq
\gamma^{-1}$, where the Universe appears to be open. This scenario
was recently discussed in Ref.~\cite{quasi}.
Here we will give a brief account of the main results.

After tunneling in the $\sigma$-field direction, there remains in the
$\phi$-field (inflaton) direction a semiclassical displacement
characterized by a Gaussian distribution\footnote{Here, and for the
rest of the paper, we make the assumption that the slow-roll field
does not affect the geometry outside the bubble (which we take to be
de Sitter space). If the slow roll field has a mass outside the
bubble, then its quantum fluctuations will drive it to large values
where its potential energy substantially modifies the local expansion
rate.  This may have an effect on the distribution of the field inside
the bubble, which requires further investigation.}  with
r.m.s. amplitude $f$, where
\begin{equation}
f^2 \equiv \langle\phi^2(t_0)\rangle \approx {H_F^2 \over (2\pi)^2 \gamma}
\label{f}
\end{equation}
If $f$ is sufficiently large (for a power law potential this means
$f\sim M_{\rm Pl}$, whereas for the open hybrid model~\cite{GBL}, a much
smaller $f$ would do), then the fluctuations of the field $\phi$ will
make inflation generic inside the bubble. However, regions of size
$r\sim \gamma^{-1}$ with large positive $\phi$ will be separated from
regions with large negative $\phi$ by non-inflating regions where $\phi$
is small.

In many models, however, the r.m.s. fluctuation $f$ is much smaller than
the field value needed for inflation. Then, most of the hypersurface
$t=t_0\sim H^{-1}$ inside the bubble will not be inflating, leading to
empty space with no galaxies. On the other hand, inflating regions will
still arise as localized ``rare'' fluctuations, with exponentially
suppressed probability
\begin{equation}\label{prob}
P\propto \exp(-\phi^2/2f^2),
\label{suppression}
\end{equation}
where $\phi \sim M_{\rm Pl}$. High peaks of a random Gaussian field tend
to be spherical. If we choose the origin of coordinates on the $t=const$
hyperboloid to be at the center of the island, then the profile of the
field as we move outwards is given by the $l=0$ supercurvature mode [see
Equation (A4)], normalized to the value $\phi_c$ at the center of the
bubble~\cite{quasi},
\begin{equation}
\phi^{\gamma}_{l=0} = \phi_c(t) {\sinh(\Lambda r) \over \Lambda \sinh r}.
\label{l=0}
\end{equation}
The r.m.s. amplitude of the remaining $l>0$ supercurvature modes, which
would account for departures from sphericity, is much smaller, of
order $\gamma^{1/2} f \ll f \ll \phi(t_0)$. 

Let us concentrate on one of these inflating regions. The initial value
of $\phi_c$ determines how many e-foldings of inflation the center of
the island will undergo, and hence the value of $\Omega_0$ that an
observer in that region would measure after inflation, at a given CMB
temperature.  For the sake of illustration, let us assume that this
observer measures $\Omega_0 = 0.5$ at the time when $T_{\rm
CMB}=2.728$~K.  Also, let us take $\gamma=10^{-4}$. For $r\ll
\gamma^{-1}\sim 10^4$ the field on a $t=const$ slice will decrease very
slowly with distance as we move away from the center. Note that at large
distances from the center $r\gg1$ we have $\phi^{\gamma} \propto
\exp(-\gamma r/2)$. Denoting by $N$ the number of e-foldings of
inflation and assuming a quadratic potential (a similar argument can be
made for the hybrid model), we have $\delta N/N \approx 2
\delta\phi/\phi$. Using the relation $2 \delta
N=[\Omega_0(\Omega_0-1)]^{-1}\delta \Omega_0$, we find that observers
out to a distance $r\sim 10$ would measure a very similar density
parameter, which differs from the one at the center only by
$\delta\Omega_0\approx \Omega_0(\Omega_0-1)N\gamma r \lesssim 10^{-2}$.
On the other hand, for $r\sim \gamma^{-1} = 10^4$ the universe will look
rather empty, and even if inflation proceeds there for a few e-foldings,
the density parameter would be too low for any galaxies to form and for
observers to develop.

Although the inflating region in the example above has spherical
symmetry around $r=0$, it is clear that most observers in that island
will live at $r \gg 1$. To them the universe would look
anisotropic. This effect was dubbed ``classical
anisotropy''~\cite{slava}, and can be estimated as follows.  After
expansion of (\ref{l=0}) for $r\ll\gamma^{-1}$, we can separate the
$t$-dependent background from the $t$ and $r$ dependent perturbation,
$\phi^{\gamma}_{l=0}=\phi_c(t)+\delta\phi$, where
\begin{equation}\label{deltaphi}
\delta\phi=\phi_c(t){\gamma\over 8}(1-4 r\coth r)\approx
-\phi_c(t){\gamma\over2}\ln\cosh r\,,
\end{equation}
the last expression being a very good approximation for $1\ll
r\ll\gamma^{-1}$. To describe the Universe from the point of view of an
observer living at $r=r_0\gg 1$, it is convenient to change the
coordinates $(r,\theta,\phi)$ on the spacelike hyperboloid to a new set
$(r',\theta',\phi')$ such that the point $r=r_0$ is now the new origin
of coordinates, $r'=0$. In that case, we have $\cosh r=\sinh r_0\sinh r'
\cos\theta' + \cosh r_0 \cosh r'$, and the field can be separated in the
form $\phi^{\gamma}_{l=0}=\phi_0(t)+\delta_0\phi$, where $\phi_0\approx
\phi_c(1-\gamma r_0/2)$ is the value of the field at the location $r_0$,
which corresponds to a somewhat lower value of $\Omega_0$ than that
of the central region, and
\begin{equation}
\delta_0\phi\approx\phi_0(t)(\gamma/2)\ln f(r',\theta'), 
\label{pertd}
\end{equation}
with
\begin{equation}\label{frtheta}
f(r,\theta)\equiv \cosh r + \sinh r\,\cos\theta\,.
\end{equation}

We can now evaluate the gauge invariant metric perturbation associated 
with this field fluctuation, which will remain constant
outside the horizon and will reenter during the matter era with an
amplitude
\begin{equation}\label{PHI}
\Phi = {3\over5}\,{H_T\delta\phi\over\dot\phi_0} =
{3\over5}\,A_C\,\ln f(r,\theta)\,,
\end{equation}
where
\begin{equation}\label{AC}
A_C = {3\over2}{H_T^2\over m_T^2}\,\gamma \,; \ \ 
\gamma = {2\over3}{m_F^2\over H_F^2} + {1\over8}H_F^2R_0^4
(m_T^2-m_F^2)\,.
\end{equation}
Here $R_0$ is the radius of the bubble at tunneling~(\ref{RHT}), $m_F$
and $m_T$ are respectively the masses of the inflaton field in the false
and in the true vaccuum, and $\gamma$ was computed, to order $(H_F
R_0)^4$, in Ref.~\cite{quasi}.  Note that in the case of the simplest
(uncoupled) two-field model~\cite{LM}, where $m_T=m_F$, these
expressions coincide with those given in Ref.~\cite{slava}.

In deriving (\ref{pertd}) we have concentrated in a single island and
we have assumed that we live far from the center, at $r\gg1$. This
seems to be a minimal ``Copernican'' requirement, since in any island
there are many more observers far from the center than near it. Hence,
we should (at least) impose the constraint that the CMB anisotropy
induced by (\ref{PHI}) should not exceed the observational bounds.

However, this is not the full story \cite{gtv}. Let us note, first of
all, that the amplitude of the perturbation (\ref{pertd}) is
proportional to $\phi_0$. Now, when the ensemble of all possible
islands that contain a particular value of $\phi_0$ is considered, one
finds that those values of $\phi_0$ for which (\ref{pertd}) would be
larger than the perturbation caused by the usual supercurvature modes
will occur typically near the center of the islands (see Appendix B).
Hence, those are rather unlikely values of $\phi_0$ for us to observe.
Therefore it seems reasonable to impose not only that the anisotropy
(\ref{PHI}) should not exceed the observational bounds, but also that
it should not exceed the anisotropy created by the $l>0$
supercurvature modes.

These considerations naturally lead to the question of what are the
most likely values for $\phi_0$ or what is the most probable value of
$\Omega_0$ in a given model. The probability distribution for
$\Omega_0$ in models of quasiopen inflation was studied in
Ref.~\cite{gtv}.  Stronger constraints on the models can be obtained
from this probability distribution in combination with CMB
constraints.  These will be discussed in Section VII. 

\section{CMB temperature anisotropies}

Quantum fluctuations of the inflaton field $\phi$ during inflation
produce long-wavelength scalar curvature perturbations and tensor
(gravitational waves) perturbations, which may leave their signature
in the CMB temperature aniso\-tropies, when they re-enter the
horizon. Temperature anisotropies are usually given in terms of the
two-point correlation function or power spectrum $C_l$, defined by an
expansion in multipole number $l$.  We are mainly interested in the
large-scale (low multipole number) temperature anisotropies since it
is there that gravitational waves and the discrete modes could become
important. After $l\sim50$, the tensor power spectrum drops
down~\cite{Starobinsky,HW} while the density perturbation spectrum
increases towards the first acoustic peak, see Ref.~\cite{Silk}. On
these large scales the dominant effect is gravitational redshift via
the Sachs--Wolfe effect~\cite{SW67}. 

\subsection{Scalar and tensor anisotropies}\label{scaten}

The Sachs-Wolfe effect for scalar perturbations with
adiabatic initial conditions, like those present here, is given
by~\cite{SW67}
\begin{eqnarray}\label{dTT}
{\delta T\over T}(\theta,\phi) &=& 
{1\over3}\Phi(0)Q(\eta_0,\theta,\phi) \nonumber\\
&+& 2\int_0^{\eta_0}dr\,\Phi'(\eta_0-r)Q(r,\theta,\phi)\,,
\end{eqnarray}
where $\eta_0=\cosh^{-1}(2/\Omega_0-1)$ is the present conformal time
and, to very good approximation, also the distance to the last
scattering surface ($\eta_{\rm LSS}\approx0$). Here $Q(r,\theta,\phi)$
stands for the spatial dependence of the fluctuation~\cite{JGB}.  The
second term accounts for integration along the line of sight, and is
important in an open universe, since the metric perturbation $\Phi$
evolves in time after reentering the horizon, $\Phi(\eta)=\Phi_{\rm
LS}\,F(\eta)$, where $\Phi_{\rm LS}$ is the value of the perturbation at
the moment of last scattering and
\begin{equation}\label{Feta}
F(\eta)\equiv5{\sinh^2\eta-3\eta\sinh\eta+4(\cosh\eta-1)\over
(\cosh\eta-1)^3}\,,
\end{equation}
normalized so that $F(0)=1$ at last scattering, $\eta_{\rm LSS}=0$. 

One can always expand the temperature anisotropy projected on the sky
in spherical harmonics,
\begin{equation}\label{Ylm}
{\delta T\over T}(\theta,\phi) = \sum_{l=1}^{\infty}\sum_{m=-l}^{l}\,
a_{lm}\,Y_{lm}(\theta,\phi)\,,
\end{equation}
where the coefficients $a_{lm}$ can be obtained from
\begin{equation}\label{alm}
a_{lm}=\int d\Omega\,Y_{lm}^*(\theta,\phi){\delta T\over T}(\theta,\phi)\,.
\end{equation}

Perturbations generated during inflation have generically Gaussian
statistics since they arise from linearized quantum fluctuations. As a
consequence, one can characterize the temperature anisotropies just
with the two-point correlation function, or angular power spectrum,
\begin{equation}\label{Cls}
\left\langle{\delta T\over T}(\hat{\bf n})\cdot
{\delta T\over T}(\hat{\bf n}')\right\rangle_{\!\!\hat{\bf n}
\cdot\hat{\bf n}'=\cos\theta} \hspace{-3mm}
= \sum_{l=1}^{\infty}\,{2l+1\over4\pi}\,C_l P_l(\cos\theta)\,,
\end{equation}
where $(2l+1)C_l = \sum_{m=-l}^{l}\langle|a_{lm}|^2\rangle$,
brackets indicating averages over different realizations.

\begin{figure}[t]
\centering
\hspace*{-5mm}
\leavevmode\epsfysize=11cm \epsfbox{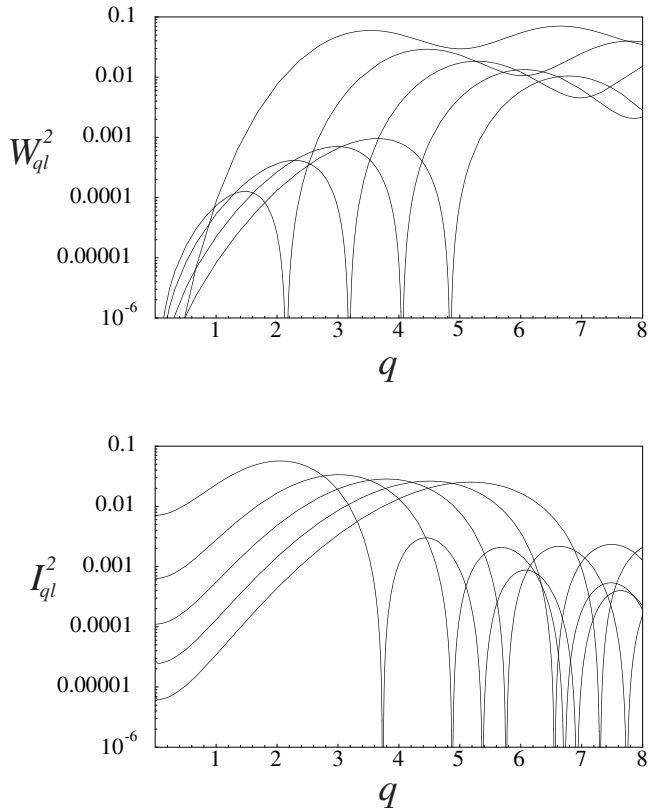}\\[3mm]
\caption[fig6]{\label{fig6} Window functions for the scalar (top figure)
  and tensor (bottom figure) CMB power spectra, for the first few
  multipoles, $l=2,4,6,8,10$ (from top-left to bottom-right), for
  $\Omega_0=0.4$. Note that the scalar window function $W_{ql}$ goes
  to $0$ as $q\to 0$, whereas the tensor window function $I_{ql}$
  remains finite in this limit. It is for this reason that the shape
  of the function $f(q)$ is very important for the tensor but not much
  for the scalar component of temperature anisotropies.}
\end{figure}

The scalar and tensor components of the temperature power spectrum can
be computed as 
\begin{eqnarray}\label{CLS}
  D_l^S \equiv l(l+1)C_l^S &=& l(l+1)\int_0^\infty dq\,
\langle|{\cal R}(q)|^2\rangle  \,W_{ql}^2\,, \\[1mm]
  D_l^T \equiv l(l+1)C_l^T &=& l(l+1)\int_0^\infty dq\,
\langle|h(q)|^2\rangle\,I_{ql}^2\,,
 \label{CLT}
\end{eqnarray}
where $W_{ql}$ and $I_{ql}$~\cite{JGB} are the corresponding window
functions~\footnote{\label{eigapp}The eigenfunctions $\Pi_{ql}$ for the
subcurvature modes, $\bar\Pi_{\Lambda,l}(r)$ for the supercurvature
($q=i\Lambda$, $0>\Lambda>1$) and bubble wall ($\Lambda=2$) modes
(below), and $G_{rr}^{ql}$ for the tensor modes can be found in the
Appendix A.},
\begin{eqnarray}
W_{ql} &=& {1\over5} \Pi_{ql}(\eta_0)+
{6\over5}\int_0^{\eta_0}dr
F'(\eta_0-r)\Pi_{ql}(r)\\
I_{ql} &=& \int_0^{\eta_0}drG'_q(\eta_0-r)G_{rr}^{ql}(r)\,,
\end{eqnarray}
which depend on the particular value of $\Omega_0$. Here, the function
$G_q(\eta)$ is given by \cite{TS,JGB} 
\begin{equation}
G_q(\eta) = 3\frac{\sinh \eta\sin q\eta - 2 q \cos q\eta (\cosh \eta-1)}
{q(1+4q^2)(\cosh\eta-1)^2}\,.
\end{equation}
We have plotted these functions in Fig.~\ref{fig6} for a typical
value, $\Omega_0=0.4$, and the first few multipoles. The scalar window
functions $W_{ql}$ grow as a large power of $q$ at the origin, so that
the scalar power spectrum is rather insensitive to the `hump' in the
function $f(q)$, as mentioned above.  On the other hand, the tensor
window functions $I_{ql}$ remain finite at $q=0$, and only the linear
dependence of $f(q)$ at the origin prevents the existence of the
infrared divergence found in \cite{Allen}.  Furthermore, since the
functions $I_{ql}^2$ are not negligible near the origin, the tensor
power spectrum turns out to be very sensitive to the `hump' in the
spectral function $f(q)$.

The ratio of tensor to scalar contribution to the temperature power
spectrum is a fundamental observable in standard inflation, which
depends on the slow-roll parameters (\ref{slowroll}) and provides a
consistency check of the theory~\cite{LL93}. In single-bubble open
inflation, such a ratio depends not only on the slow-roll parameters
but also on the tunneling parameters (\ref{ab}) and on the value of
$\Omega_0$, see Ref.~\cite{JGB}:
\begin{equation}\label{Oratio}
R_l = {C_l^T\over C_l^S} \simeq f_l(\Omega_0,a,b)\,
[1-1.3(n_S-1)]\,16\,\epsilon \,. 
\end{equation}
In the ideal case in which the gravitational wave perturbation may be
disentangled from the scalar component in future precise observations
of the CMB power spectrum~\cite{MAP,COBRAS,JKKS,ZSS,BET}, 
it might be possible to test this relation for a given value of
$\Omega_0$.  This would constitute a check on the tunneling
parameters $a$ and $b$.  Such prospects are, however, very bleak from
measurements of the temperature power spectrum alone, with the next
generation of satellites, see e.g.~\cite{Knox,COBRAS}. One can at most
expect, from the absence of a significant gravitational-wave
contribution to the CMB, to impose constraints on the parameters of the
model. However, taking into account also the polarization power
spectrum, together with the temperature data, one expects to do much
better in the case of flat models, see Refs.~\cite{SZ,ZSS}. Hopefully
similar conclusions can be reached in the context of open models, and
CMB observations may be able to check the generalized consistency
relation (\ref{Oratio}) with some accuracy.

\subsection{Supercurvature anisotropies}\label{otheranis}

\begin{figure}[t]
\centering
\hspace*{-4mm}
\leavevmode\epsfysize=11.2cm \epsfbox{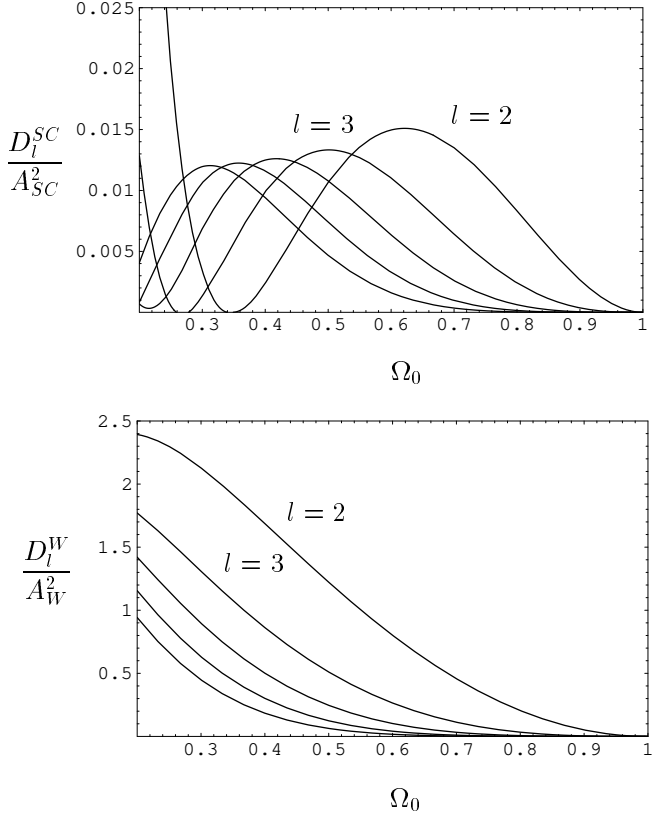}\\[3mm]
\caption[fig2]{\label{fig2} CMB power spectrum $l(l+1)\,C_l/A^2$, 
  normalized to the corresponding amplitude, for the supercurvature
  and bubble-wall fluctuations, as a function of $\Omega_0$, for the
  first few multipoles $l=2,3,...6$ Note the dip in the supercurvature
  power spectrum at different values of $\Omega_0$
  for different multipoles, due to accidental cancellations.}
\end{figure}

The supercurvature and bubble-wall modes also contribute to the
temperature power spectrum.  The supercurvature mode's contribution to
the CMB anisotropies can be written as
\begin{equation}\label{DLSC}
D_l^{SC}\equiv l(l+1) C_l^{SC} = \pi^2 A^2_{SC}\,l(l+1)\bar W_{\Lambda,l}^2\,,
\end{equation}
with the window function
\begin{equation}\label{Wql}
\bar W_{\Lambda,l} = {1\over5}\bar \Pi_{\Lambda,l}(\eta_0)+
{6\over5}\int_0^{\eta_0}dr
F'(\eta_0-r)\bar\Pi_{\Lambda,l}(r)\,,
\end{equation}
and the amplitude $A^2_{SC}$ is given by eq. (\ref{ASC})

The bubble wall mode's contribution to the CMB anisotropies can be
written as
\begin{equation}\label{DLW}
D_l^W\equiv l(l+1) C_l^W = 4\pi^2 A^2_W\,l(l+1)\bar W_{2,l}^2\,,
\end{equation}
where the window function is given by Eq.~(\ref{Wql}) for $\Lambda=2$,
and the amplitude $A^2_W$ is found in Eq. (\ref{A2W})

Their contribution is plotted in Figs.~\ref{fig2} and~\ref{fig1},
normalized to the corresponding amplitude, see Eqs.~(\ref{ASC}),
(\ref{A2W}) and (\ref{AC}), as a function of $\Omega_0$ for the first
few multipoles.  Note the dip in the spectrum at certain values of
$\Omega_0$ due to accidental cancellations~\cite{induced,GBL,JGB}
between the intrinsic and integrated Sachs-Wolfe effects. This does not
affect the bounds, since higher multipoles will fill in those gaps. The
relative importance of the different components of the power spectrum is
crucial in order to derive bounds on the model parameters.  We have
plotted in Fig.~\ref{fig3} the quadrupole and tenth multipole of the CMB
power spectrum for each mode, normalized to their corresponding
amplitudes.

\subsection{Classical anisotropies}

We will concentrate here on the anisotropies associated with
quasiopenness, the classical temperature anisotropies. In that case,
Eq.~(\ref{dTT}) can be written as
\begin{eqnarray}\label{dT}
{\delta T\over T}(\theta) &=& {1\over5}A_C\ln f(\eta_0,\theta) 
\nonumber \\
&+&{6\over5}\int_0^{\eta_0}dr\,F'(\eta_0-r)\,A_C\ln f(r,\theta)\,.
\end{eqnarray}

\begin{figure}[t]
\centering
\hspace*{-4mm}
\leavevmode\epsfysize=10.8cm \epsfbox{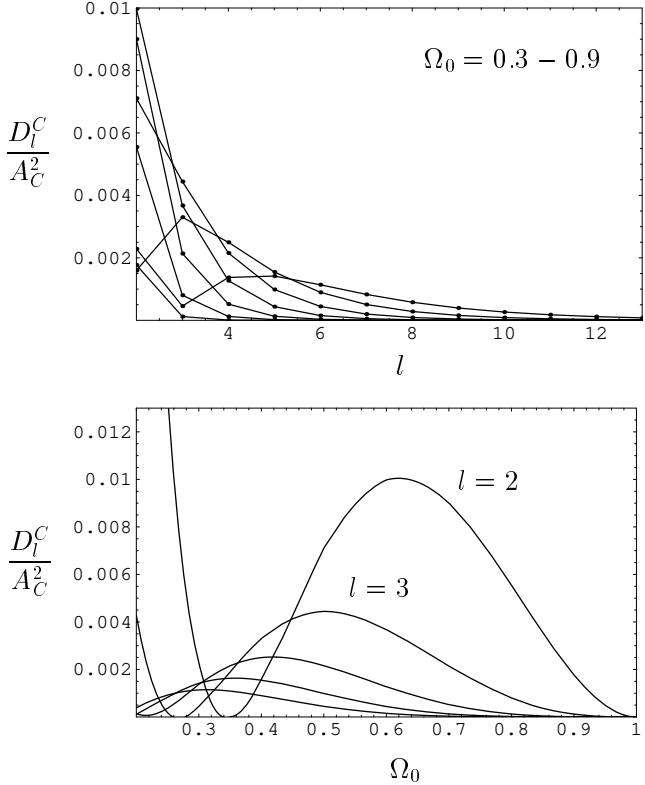}\\[3mm]
\caption[fig1]{\label{fig1} CMB power spectrum $l(l+1)\,C_l^C$,
  normalized to the corresponding amplitude $A_C^2$, for the
  semi-classical fluctuations, as a function of $\Omega_0$, for the
  first few multipoles $l=2,3,...7$, and as a function of multipole
  number $l$, for $\Omega_0=0.3-0.9$. Note the dip in the
  semi-classical power spectrum at different values of $\Omega_0$ for
  different multipoles, due to accidental cancellations.}
\end{figure}

Expanding in spherical harmonics (\ref{Ylm}), we can write $a_{lm}$ as
(\ref{alm}). In our case, since the temperature anisotropy (\ref{dT})
only depends on $\theta$, we find
\begin{eqnarray}\label{al0}
a_{lm}&=&\left[{2l+1\over4\pi}\right]^{1/2}\!2\pi\!
\int\!d\theta \,\sin\theta \,P_l(\cos\theta)\,
{\delta T\over T}(\theta)\ \delta_{m0} \\
&\equiv& a_{l0}\,\delta_{m0}\,.
\end{eqnarray}

It implies that the coefficients $C_l$ satisfy, in this case,
\begin{equation}\label{ClC}
(2l+1)C_l^C = \sum_{m=-l}^{l}\langle |a_{lm}|^2\rangle = 
\langle |a_{l0}|^2\rangle\,.
\end{equation}
Therefore, we find that the power spectrum associated with the
classical anisotropies (\ref{PHI}) can be computed as
\begin{eqnarray}\label{Dl}
D_l^C&\equiv& l(l+1)C_l^C = \pi\,A_C^2 \,l(l+1) W_l^2 \,,\\
W_l &\equiv& \int_0^\pi d\theta\,\sin\theta\,P_l(\cos\theta)
\Big[{1\over5}\ln f(\eta_0,\theta) \nonumber \\
&+& {6\over5}\int_0^{\eta_0}dr\,F'(\eta_0-r)\ln f(r,\theta)\Big]\,.
\end{eqnarray}

We have plotted this expression in Fig.~\ref{fig1}, as a function of
multipole number $l$, for various values of $\Omega_0$; and as a
function of $\Omega_0$, for the first few multipoles. As can be
appreciated from the figure, the window functions for the
semiclassical mode for a given $l$ are proportional to the
corresponding ones for the supercurvature modes. This was
expected since, as mentioned in Section IV and in Appendix B, the
quasiopen island can be thought of as a superposition of
supercurvature modes.

\subsection{Bounds from CMB anisotropies}

\begin{figure}[t]
\centering
\hspace*{-4mm}
\leavevmode\epsfysize=11.5cm \epsfbox{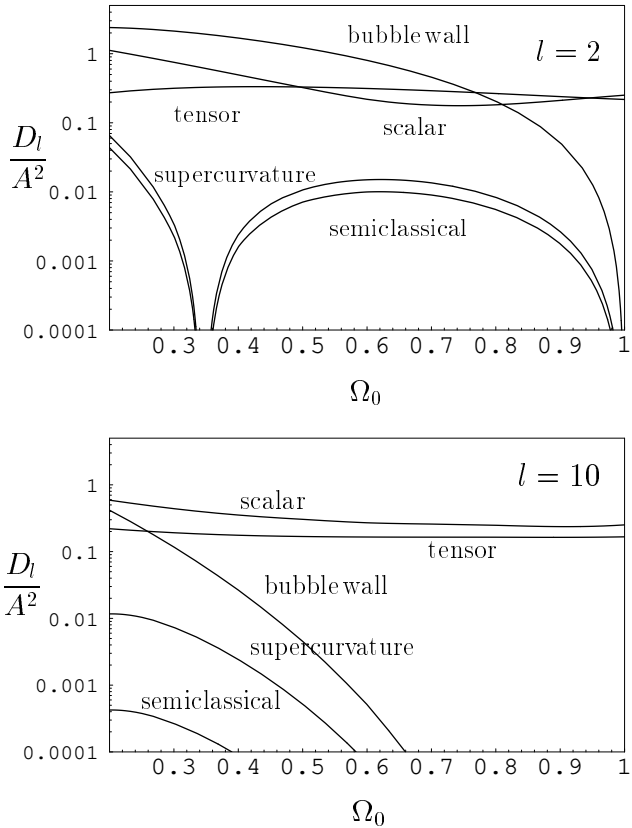}\\
\caption[fig3]{\label{fig3} Quadrupole (top figure) and tenth multipole
  (bottom figure) of the CMB power spectra, normalized to the
  corresponding amplitude, $l(l+1)\,C_l/A^2$, for the tensor, scalar,
  supercurvature, bubble-wall and semi-classical primordial spectra.
  We are assuming here the minimal contribution from tensors 
  ($a=0, b=1$).}
\end{figure}

\begin{figure}[t]
\centering
\hspace*{-4mm}
\leavevmode\epsfysize=5cm \epsfbox{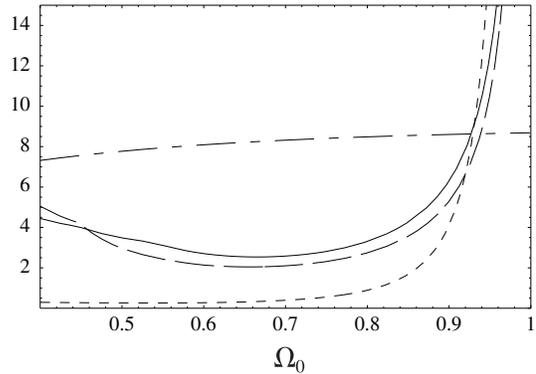}\\[2mm]
\caption[fig7]{\label{fig7} This figure shows the bounds
  (\ref{bounds})--(\ref{gamma}) as a function of $\Omega_0$. The
  dotted-dashed line corresponds to the actual constraint on the
  scalar component, $(H_T/\sqrt\epsilon M_{\rm Pl})\times10^5$ for a
  scale invariant scalar spectrum. The rest are upper bounds, on
  $H_F/H_T$ (solid line), on $A_C\times10^4$ (dashed line), and on
  $\epsilon/z$ (dotted line). As can be seen, for $\Omega_0\simeq1$,
  the upper bounds are not very restrictive.}
\end{figure}

From the 4-year Cosmic Background Explorer (COBE) maps~\cite{COBE},
the overall amplitude and tilt of the CMB temperature power spectrum
at small multipole number have been determined with some accuracy for
$\Omega_0\simeq1$ \cite{Bond}
\begin{eqnarray}\label{Bond}
\left[{l(l+1)C_l\over2\pi}\right]^{1/2}\!&=&
(1.03\pm0.07)\times 10^{-5}\,,\\ 
n &=& 1.02\pm0.24\,.
\end{eqnarray}
For an open Universe, Bunn and White gave a compact expression~\cite{BW}
\begin{eqnarray}\label{BW}
A_S&=&4.9\times10^{-5}\,{\Omega_0\over g(\Omega_0)}\,\nonumber
\Omega_0^{-0.35 - 0.19 \ln\Omega_0 - 0.17(n_S-1)}\\
&&\times\exp[-(n_S-1)-0.14(n_S-1)^2]\,,
\end{eqnarray}
where $g(\Omega)/\Omega = 5/2(1+\Omega/2+\Omega^{4/7})^{-1}$ is a
fitting function to the suppression in the growth of scalar
perturbations in an open universe relative to the critical density
universe \cite{lial95}, and it was assumed that only the scalar
component contributed significantly to the CMB anisotropies. Under this
assumption, one can deduce the following constraints, for a scale
invariant spectrum, see Refs.~\cite{JGB,quasi},
\begin{eqnarray}\label{bounds}
{H_T\over\sqrt\epsilon M_{\rm Pl}}&=&\sqrt\pi A_S
\approx 9\times 10^{-5}\,, \\  \label{super} 
{H_F\over H_T} &<& \sqrt{\frac{D_l^S/A_S^2}{2D_l^{SC}/A_{SC}^2}} 
\approx 3\,,\\ \label{bubble} 
{\epsilon\over z} &<& \frac{D_l^S/A_S^2}{D_l^W/A_W^2}
\approx 0.6\,,\\ \label{gamma} 
A_C &<& \sqrt{\frac{D_l^S/A_S^2}{D_l^C/A_C^2}}\,A_S { } 
\approx 3\times 10^{-4}\,,
\end{eqnarray}
where the approximations are valid through the range $0.4\lesssim
\Omega_0 \lesssim0.9$, for the quadrupole, see Fig.~\ref{fig7}. The
third expression accounts for both the tensor and the bubble-wall
constraints, since the bubble-wall fluctuation is actually part of the
tensor spectrum~\cite{new} and gives the largest contribution at low
multipoles. The constraints coming from higher multipoles are
significantly weaker, see Fig.~\ref{fig3}. One could argue that the
quadrupole is going to be hidden in the cosmic variance~\cite{BET} and
thus only the constraints at higher multipoles, say $l>2$, should be
imposed.  However, polarization power spectra may one day be used to
get around cosmic variance~\cite{Loeb} and we may be able to extract
information about the scalar and tensor components at low multipoles.
We will therefore take the conservative attitude that consistent
models of open inflation should satisfy the bounds coming from the
first few multipoles, as long as they do not exceed the associated
cosmic variance.

The bounds (\ref{bounds})-(\ref{gamma}) are $\Omega_0$-dependent. For
values of $(1-\Omega_0)\ll1$ analytic expressions for the temperature
anisotropy can be found as a power series in $(1-\Omega_0)$. In the
limit $\Omega_0\to1$, the scalar and the tensor contribution remain
finite, whereas the supercurvature and the semiclassical anisotropies
vanish, as can be seen from Fig.~\ref{fig3}. Due to this fact, in the
limit $\Omega_0\to 1$, the bounds (\ref{super})-(\ref{gamma})
disappear and put no constraint on the parameters of the models.  The
scalar contribution in this limit is found to be given by\cite{JGB}
\begin{equation}\label{cls}
C_l^S = {2\pi A^2_S\over25}{\Gamma(3/2)\Gamma((3-n_S)/2)
\Gamma(l+(n_S-1)/2)\over\Gamma(2-n_S/2)\Gamma(l+2-(n_S-1)/2)}\,.
\end{equation}
Hereafter we will restrict ourselves to the scale invariant case,
$n_S=1$. Then equation (\ref{cls}) reduces to the well known result
$D_l^S=2\pi A_S^2/25$. In the limit $\Omega_0\to 1$, the tensor
contribution can be expressed as~\cite{Starobinsky}
\begin{equation}
D_l^T = {\pi A_T^2\over36}\Big(1+{48\pi^2\over385}\Big)\,B_l\,,
\end{equation}
where $B_l = (1.1184, 0.8789, \dots)$, for $l = 2, 3, \dots$, which
approaches $B_l = 1$ for large multipoles ($l\sim10$) and then
decays to zero after $l\sim50$.

For the supercurvature and semiclassical temperature anisotropies,
expanding $\bar \Pi_{ql}(\eta_0)$ and $f(\eta_0,\theta)$ in powers of
$\eta_0\approx 2(1-\Omega_0)^{1/2}$, it is easily seen that the
contribution to the corresponding window functions of the Integrated
Sachs-Wolfe effect is suppressed by an extra power of $(1-\Omega_0)$
with respect to the intrinsic Sachs-Wolfe contribution. Taking this
into account, the leading contribution to the temperature anisotropy
is given by the following expressions:
\begin{eqnarray}
D^{SC}_l &=&
\frac{\pi^2 A_{SC}^2}{100}\frac{\Gamma(l+2)^2}{\Gamma(l+3/2)^2}
(1-\Omega_0)^l\,,\\
D^{W}_l &=&
\frac{\pi^2 A_{W}^2}{25}\frac{\Gamma(l+2)\Gamma(l+3)}{(l-1)
\Gamma(l+3/2)^2}(1-\Omega_0)^l\,,\\
D^{C}_l &=& \frac{\pi^2 A_C^2}{25}\frac{\Gamma(l+2)\Gamma(l+1)}
{l\,\Gamma(l+3/2)^2}(1-\Omega_0)^l\,. 
\end{eqnarray}
Better approximations can be found by just including more terms in the
expansions of $\bar\Pi_{ql}(\eta_0)$ and $f(\eta_0,\theta)$, but this
will suffice for our purposes. 

Then, for values of $(1-\Omega_0)\ll 1$, the bounds read
\begin{eqnarray}
\label{bsuperapp}
{H_F\over H_T} &\lesssim& {0.625\over(1-\Omega_0)}\\
\label{bwallapp}
{\epsilon\over z} &\lesssim&
\ {0.044\over(1-\Omega_0)^2}\\
\label{bclassapp}
A_C & \lesssim & {5.3\times10^{-5}\over(1-\Omega_0)}\,.
\end{eqnarray}
The error commited using this approximation is around $5\%$
for the supercurvature and semiclassical bounds for values of
$\Omega_0\approx0.8$, and less than $10\%$ for the bubble wall
bound and values of $\Omega_0\approx0.9$.
 
\section{Model building}

In this section we will review the different single-bubble open
inflation models present in the literature, and use the CMB
observations to rule out some of them and severely constrain others.
We will see that open inflation models could be as predictive as
ordinary inflation, in the sense that they also can be ruled out
if they are in conflict with observations.

\subsection{single-field models}

As mentioned in Section 2, single-field models of open
inflation~\cite{singlebubble} require some fine-tuning in order to have
a large mass for successful tunneling and a small mass for slow-roll
inside the bubble~\cite{LM}. Even if such a model can be constructed
from particle physics, it still needs to satisfy the constraints coming
from observations of the CMB anisotropies. In the thin wall regime,
these models lack both supercurvature modes ~\cite{YST} and
semi-classical anisotropies, by construction.  However, in some cases,
they produce too large tensor anisotropies at low multipoles~\cite{JGB},
where they are dominated by the bubble-wall fluctuations.

Let us analyse a typical example, which is a variant of the new
inflation type~\cite{singlebubble}. The potential is a quartic double
well, to which a double barrier has been added near the
origin. Tunneling occurs from a symmetric phase at $\sigma=0$ to a
value $\sigma_b$ from which the field slowly rolls down the potential
towards the symmetry breaking phase at $\sigma=v\sim M_{\rm GUT}\sim
10^{15}$ GeV. The fact that we have a finite number of e-folds,
$N_e=60$, requires $\sigma_b\sim v\,\exp(-\alpha N_e)\ll v$, where
$\alpha\simeq m_T^2/3H_T^2$.  The rate of expansion in the true vacuum
is of the order of that in the false vacuum, $H_T\simeq (8\pi
V(0)/3M_{\rm Pl}^2)^{1/2} \sim 3\times10^{-6} M_{\rm Pl}$, which implies
$\epsilon\sim10^{-3}$, for agreement with CMB anisotropies. Choosing a
typical mass in the false vacuum to be $M\sim M_{\rm GUT}$, we find
\begin{equation}
b_{\rm single} \simeq \Big({\sigma_b\over M_{\rm Pl}}\Big)^2 
{M\over H_T} \sim 5\times10^{-9}\,,
\end{equation}
which gives $z\sim 2b \sim 10^{-8}$, an extremely small number
that makes it impossible to satisfy the bubble-wall constraint,
$\epsilon < z$, see Eq.~(\ref{bubble}). 
In other words, the simplest single-field models of open
inflation~\cite{singlebubble} are not only fine-tuned but actually
produce too large gravitational-wave anisotropies in the CMB on large
scales to be consistent with observations. The reason for that is
that in a new inflation type potential, slow roll has to begin very
close to $\sigma=0$ in order to have sufficient inflation. However,
this does not leave much room for a sufficiently thick barrier.

Linde has recently proposed a new single-field open inflation
model~\cite{newsingle} in which the two different mass scales needed for
tunneling and for slow-roll can coexist. This is basically a quadratic
potential, where a barrier is {\em appended} at $\phi\sim 3 M_{\rm
Pl}$. Although the model is somewhat {\em ad hoc} from the point of view
of particle physics, there is in principle the possibility of making the
barrier sufficiently thick, so that the bubble wall fluctuations will
not be important. This model has specific signatures of its
own~\cite{newsingle,prep}.

\subsection{Coupled and uncoupled two-field models}
\label{twofields}
\begin{figure}[t]
\centering
\hspace*{-4mm}
\leavevmode\epsfysize=5.5cm \epsfbox{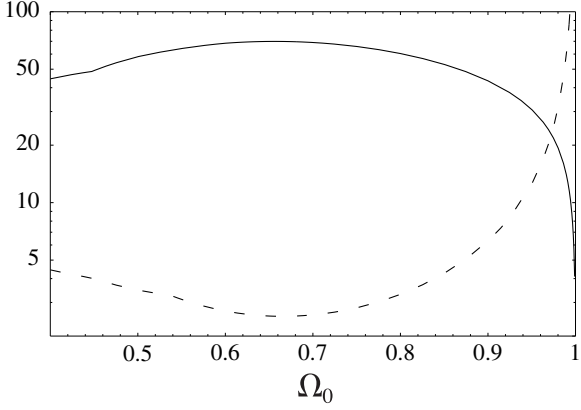}\\[2mm]
\caption[fig9]{\label{fig9} Constraints on $H_F/H_T$ due to
  supercurvature and semiclassical fluctuations in the uncoupled
  open inflation model. The region above the dashed line (due to
  supercurvature fluctuations) and below the solid line (due to
  semiclassical fluctuations) is excluded by observations. The region
  allowed by observations (the small corner to the right of the
  picture) leaves only values of $\Omega_0$ that are very close to~1.}
\end{figure}

In this section we shall consider a class of two-field
models~\cite{LM} with a potential of the form
\begin{equation}
V(\sigma, \phi)= V_0(\sigma)+{1\over 2}m^2\phi^2 + 
{1\over 2} g^2 \sigma^2\phi^2.
\label{general}
\end{equation}
Here $V_0$ is a non-degenerate double well potential, with a false
vacuum at $\sigma=0$ and a true vacuum at $\sigma=v$.  When $\sigma$
is in the false vacuum, $V_0$ dominates the energy density and we have
an initial de Sitter phase with expansion rate given by $H_F^2\approx
8\pi V_0(0)/3M_{\rm Pl}^2$. Once a bubble of true vacuum $\sigma=v$
forms, the energy density of the slow-roll field $\phi$ may drive a
second period of inflation. However, as pointed out in Ref.~\cite{LM},
the simplest two-field model of open inflation, given by
(\ref{general}) with $g=0$ and $m\neq 0$, i.e the {\sl uncoupled}
two-field model, is actually a quasi-open one; this is so because
equal-time hypersurfaces, defined by the $\sigma$ field after
nucleation, are not synchronized with equal-density hypersurfaces,
determined by the slow-roll of the $\phi$ field during inflation
inside the bubble. In order to suppress this effect it was
argued~\cite{LM} that a large rate of expansion in the false vacuum
with respect to the true vacuum, $H_F\gg H_T$, could prevent the
$\phi$ field from rolling outside the bubble and distorting the
equal-density hypersurfaces inside it.  However, this would
induce~\cite{super} a large supercurvature mode anisotropy in the CMB,
which would be incompatible with observations.  A careful
analysis~\cite{slava,quasi} shows that indeed the effect is important
at low multipoles. In this model, $m_F=m_T=m$, the supercurvature
eigenvalue $\gamma = 2m_F^2/3H_F^2$, and in order to satisfy
(\ref{gamma}) it requires 
\begin{equation}
{H_F\over H_T} > \Big({D_l^C/A_C^2\over D_l^S/A_S^2}\Big)^{1/4}
{1\over\sqrt{A_S}}\approx 60\,,
\end{equation}
which is incompatible with
the supercurvature constraint 
\begin{equation}
{H_F\over H_T} < \Big({D_l^S/A_S^2\over D_l^{SC}/A_{SC}^2}\Big)^{1/2}
{1\over\sqrt2}\approx 3\,.
\end{equation}
On the other hand, for values of $\Omega_0\gtrsim0.9$, combining the
bounds (\ref{bsuperapp}) and (\ref{bclassapp}) we find that
\begin{equation}
137\,(1-\Omega_0)^{1/2}\lesssim\frac{H_F}{H_T}\lesssim0.6\,
(1-\Omega_0)^{-1},
\end{equation}
which cannot be satisfied unless $\Omega_0\gtrsim0.97$, see
Fig.~\ref{fig9}. Therefore, this model seems to be ruled out for
$\Omega_0\lesssim0.97$. 

In order to construct a truly open model, Linde and Mezhlumian
suggested taking $m=0$ and $g\neq 0$, i.e. the {\sl coupled} two-field
model~\cite{LM}. In this way, the mass of the slow-roll field vanishes
in the false vacuum, and it would appear that the problem of classical
evolution outside the bubble is circumvented.  However, as we showed
in Ref.~\cite{quasi}, this is not exactly so, and actually the whole
class of models (\ref{general}) leads to quasi-open Universes, which
are constrained by CMB observations. Let us work out those constraints
in detail here. We will assume a tunneling potential like~\cite{LM}
\begin{equation}\label{tunpot}
V_0(\sigma)= V_0+{1\over 2}M^2\sigma^2 - \alpha M\sigma^3 +
{1\over 4} \lambda \sigma^4\,.
\end{equation}

\begin{figure}[t]
\centering
\leavevmode\epsfysize=6cm \epsfxsize=8.5cm\epsfbox{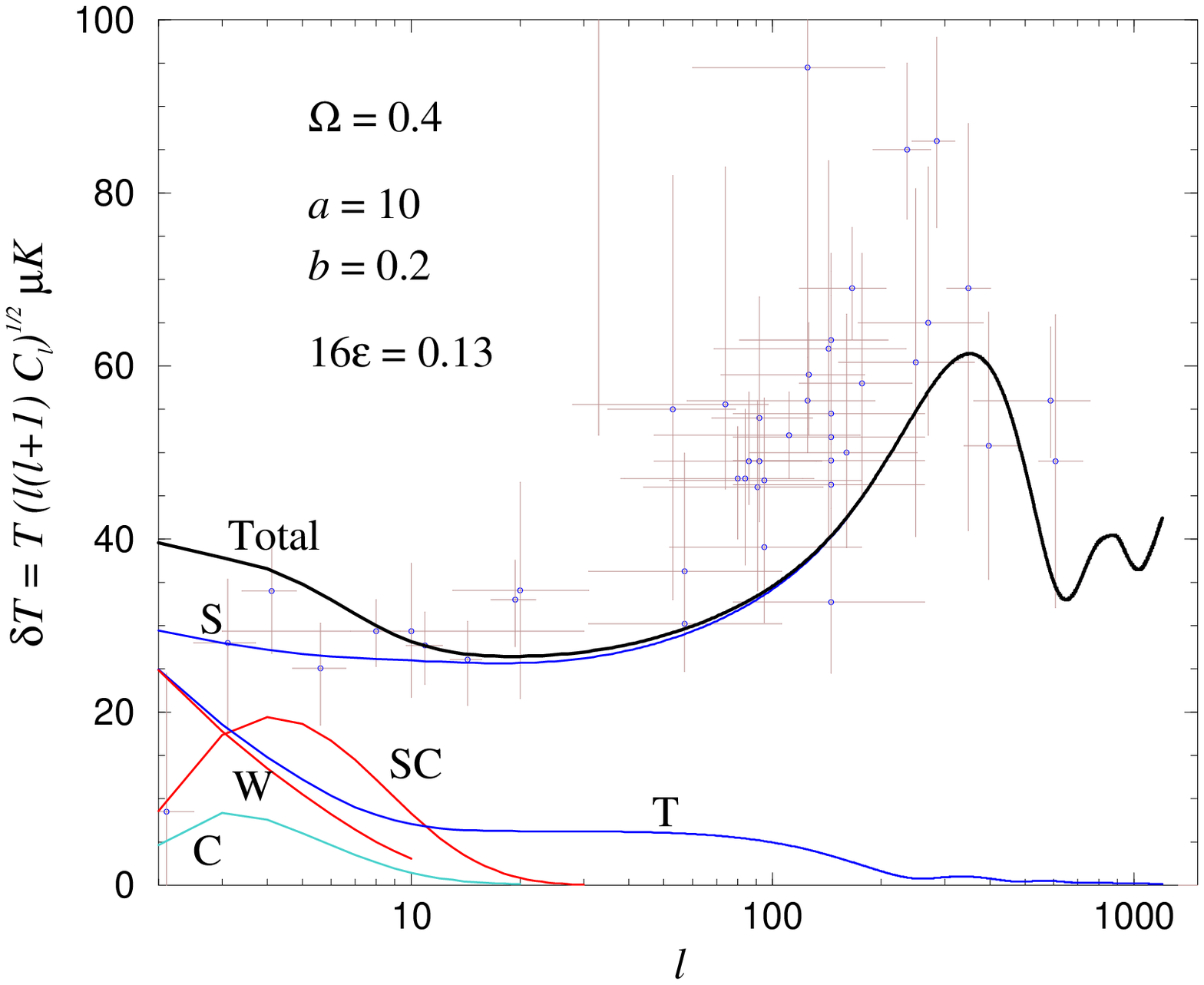}\\
\leavevmode\epsfysize=6cm \epsfxsize=8.5cm\epsfbox{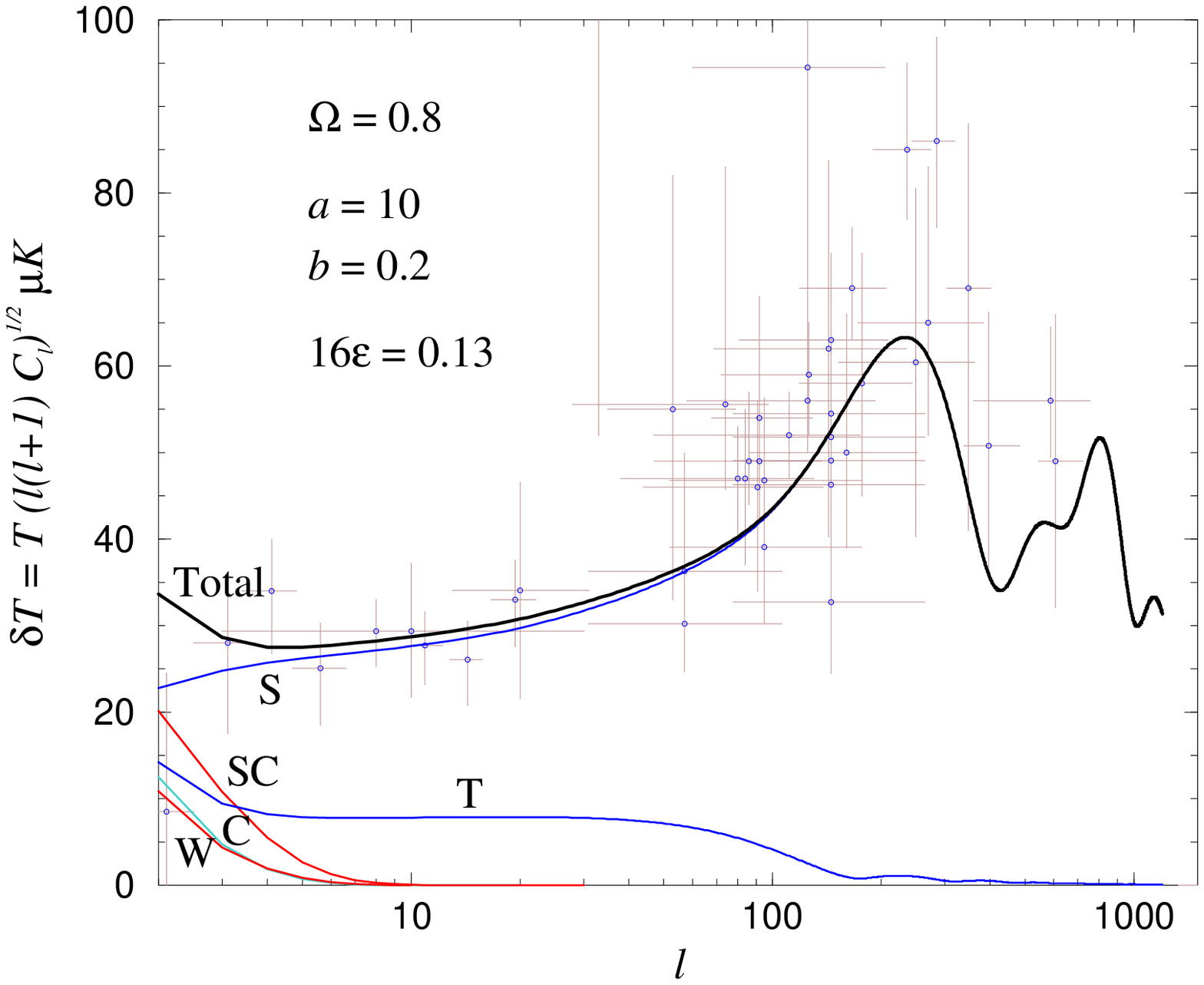}\\[3mm]
\caption[fig5]{\label{fig5} The complete angular power spectrum of
  temperature anisotropies for the coupled two-field model with
  $\Omega_0 = 0.4$ (higher plot) and $\Omega_0 = 0.8$ (lower plot), for
  $(a=10, b=0.2)$. We have chosen as cosmological parameters $h=0.70,\
  \Omega_B=0.05,\ \Omega_\Lambda=0,\ N_\nu = 3.04$.  We show the
  individual contributions from the scalar (S), tensor (T),
  supercurvature (SC), semi-classical (C) and bubble-wall (W)
  modes. Note that the bubble-wall mode is responsible for the large
  growth of the tensor contribution at low multipoles. Only the scalar
  modes remain beyond about $l=50$, where they grow towards the first
  acoustic peak. For comparison, we have superimposed the available CMB
  anisotropy data, as compiled by Tegmark \cite{tegmark}}
\end{figure}

For simplicity we will take $\alpha=\sqrt\lambda$. Then the field can
tunnel for $\phi<\phi_c=M/g$, when the minimum of the potential at
$\sigma\neq 0$ is deeper than the minimum at $\sigma=0$. The constant
$V_0\simeq2.77 M^4/\lambda$ has been added to ensure that the absolute
minimum, at $\phi=0$ and $\sigma_0\simeq1.3\,\sigma_c$, has vanishing
cosmological constant.  Here $\sigma_c$ is the minimum for
$\phi=\phi_c$. After tunneling, the field $\phi$ moves along an
effective potential $V(\phi)=m_T^2\phi^2/2$, where the effective mass
varies only slightly from tunneling to the end of inflation,
$m_T\simeq1.3\,g\sigma_c$.  This potential drives a period of chaotic
inflation with slow-roll parameters $\epsilon=\eta=1/2N_e\simeq1/120$.
Substituting into (\ref{bounds}) we find $H_T = 6.3\,m_T \simeq8\times
10^{-6} M_{\rm Pl}$, and therefore $g\sigma_c = 10^{-6} M_{\rm Pl}$.
The rate of expansion in the false vacuum is determined from
$H_F^2/H_T^2 = 1 + 4ab$, where $b = (4\pi\sqrt2/3\lambda) M^3/H_T
M_{\rm Pl}^2$, which gives
\begin{equation}\label{MHF}
M = {(1+4ab)^{1/2}\over4b}\,H_F \geq \sqrt2 H_F\,,
\end{equation}
the last condition arising from preventing the formation of the bubble
through the Hawking--Moss instanton, see Ref.~\cite{LM}. Furthermore,
taking $m_F=0$ in the equation (\ref{AC}) for the eigenvalue $\gamma$
gives
\begin{equation}\label{ACN}
A_C = {3\over16}{H_F^2\over H_T^2} (R_0H_T)^4 =  
{3(1+4ab)\over16[1+(a+b)^2]^2} <  3\times 10^{-4}\,.
\end{equation}
From the supercurvature mode condition (\ref{super}), $(1+4ab)^{1/2} <
3$, together with (\ref{MHF}) we find the constraint $b < 1/2$. From
Eq.~(\ref{ACN}), we realize that having nearly degenerate vacua,
$a\ll1$, is not compatible with observations.  Satisfying (\ref{ACN})
would require $b\ll1$ and $a\gg1$. However, for these values of the
parameters we expect large tensor contributions, see
Refs.~\cite{JGB,Review} (unless of course $\Omega_0$ is sufficiently
close to one.) 
So there should be a compromise between the
different mode contributions.

We have shown in Fig.~\ref{fig5} the complete temperature power spectrum
for a coupled two-field model having $a=10$ and $b=0.2$, for
$\Omega_0=0.4$ and $\Omega_0=0.8$, which are consistent with
observations. It has contributions from all the modes: scalar, tensor,
supercurvature, semi-classical and bubble-wall.  Note, however, that the
bubble-wall mode is in fact included in the sharp growth of the tensor
contribution at small multipole number, as emphasized in Ref.~\cite{new}
and shown explicitly in Fig.~\ref{fig5}, and should not be counted
twice. Although it is in principle possible to construct a model
consistent with observations, the parameters of such a model are not
very natural. In order to suppress the associated semi-classical
anisotropy we had to choose special values of the parameters. As can be
seen from Fig.~\ref{fig5}, there still exists a range of parameters for
which all contributions to the CMB anisotropies are compatible with
present observations. However, future observations by MAP and
Planck surveyor will help constrain or even rule out such models.

\subsection{Supernatural open inflation}

\begin{figure}[t]
\centering
\hspace*{-4mm}
\leavevmode\epsfysize=5.5cm \epsfbox{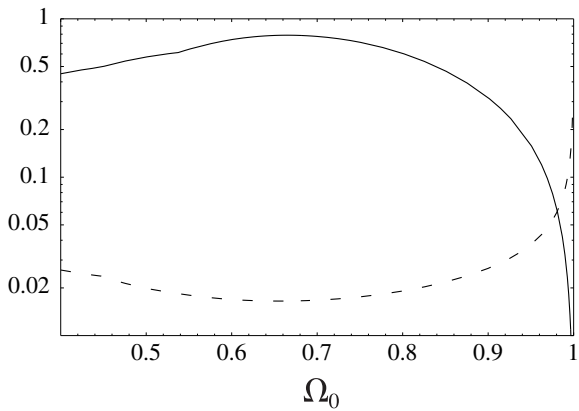}\\[2mm]
\caption[fig8]{\label{fig8} Constraints on $R_0 H_T$ due to
  supercurvature and semiclassical fluctuations in the supernatural
  open inflation model. The region below the solid line (due to
  supercurvature fluctuations) and above the dashed line (due to
  semiclassical fluctuations) is excluded by observations. The region
  allowed by observations (the small corner to the right of the
  picture) leaves only values of $\Omega_0$ that are very close to 1.}
\end{figure}

This model consists of a complex scalar field with a slightly tilted
Mexican hat potential, where the radial component of the field does the
tunneling and the pseudo-Goldstone mode does the slow-roll. This model
was called ``supernatural'' inflation in Ref.~\cite{LM}, because the
hierarchy between tunneling and slow-roll mass scales is protected by
an approximate global $U(1)$ symmetry. Expanding the field in the form
$\Phi = (\sigma/\sqrt2)\exp(i\phi/v)$, where $v$ is the expectation
value of $\sigma$ in the broken phase, we consider a potential of the
form $V=V_0(\sigma) + V_1(\sigma,\phi)$, where $V_0$ is $U(1)$-invariant
and $V_1$ is a small perturbation that breaks this invariance. It is
assumed that $V$ has a local minimum at $\Phi=0$, which makes the
symmetric phase metastable.  We shall consider a tilt in the potential
of the form $V_1= \Lambda^4(\sigma)G(\phi)$, where $\Lambda$ is a slowly
varying function of $\sigma$ that vanishes at $\sigma=0$. For
definiteness we can take $G=(1-\cos\phi/v)$. The idea is that $\sigma$
tunnels from the symmetric phase $\sigma=0$ to the broken phase
$\sigma_0=v$, landing at a certain value of $\phi$ away from the minimum
of the tilted bottom. Once in the broken phase, the potential $V_1$
cannot be neglected, and the field $\phi$ slowly rolls down to its
minimum, driving a second period of inflation inside the bubble.
Depending on the value of $\phi$ on which we end after tunneling, the
number of $e$-foldings of inflation will be different.

As in the two field model, the soft mode which corresponds to a change
in the value of $\phi$ after tunneling manifests itself as a
supercurvature mode which leads to quasiopen inflating islands.  
For the generic potential
\begin{equation}
V_1(\phi)=\Lambda^4\Big(1-\cos{\phi\over v}\Big)\,,
\end{equation}
we find the slow-roll parameters 
\begin{eqnarray}
\epsilon &=& {1\over2\kappa^2v^2}\cot^2{\phi\over2v} \ll 1\,,\\
\eta &=& \epsilon - {1\over2\kappa^2v^2} \,.
\end{eqnarray}
From the constraint on
the spectral tilt, $n_S-1=-4\epsilon-1/\kappa^2v^2 > -0.2$, we find
that, necessarily, $\kappa^2v^2 > 5$, which means that the vev of
$\sigma$ is $v\simeq M_{\rm Pl}$. We are again in a situation similar to
the single-field models, where we need some extreme fine-tuning to
prevent the Hawking--Moss instanton from forming the bubble, see
Ref.~\cite{LM}. Indeed, for a generic tunneling potential like
(\ref{tunpot}) we have $V_0\simeq M^2 \sigma_0^2/2$ and thus $H_F\simeq
2M\,\sigma_0/M_{\rm Pl}\geq M$.  Under this condition the tunneling
does {\em not} occur along the Coleman--DeLuccia instanton, which is
necessary for the formation of an open Universe inside the bubble. The
only way to prevent this is by artificially bending the potential so
that it has a large mass at the false vacuum. In Ref.~\cite{LM} a way
was proposed to lower the minimum at the center of the Mexican hat,
using radiative corrections from a coupling of the $U(1)$ field $\Phi$
to another scalar $\chi$. For certain values of the coupling constant,
$g^4=32\pi\lambda$, it is possible to make the two minima, at $\sigma=0$
and $\sigma_0$, exactly degenerate. The tunneling potential is then
\begin{equation}\label{tunpot2}
V_0(\sigma)= {\lambda\over 2}(\sigma_0^2-\sigma^2)\sigma^2 +
\lambda \sigma^4\,\ln{\sigma\over\sigma_0}\,,
\end{equation}
where $\sigma_0=M/\sqrt\lambda\simeq M_{\rm Pl}$. The associated
tunneling parameters become $a=0$ and $b=(\sigma_0/M_{\rm Pl})^2M/H_T
\simeq M/H_T$, which can be large. As emphasized in Ref.~\cite{quasi},
there is a supercurvature mode in this model, associated with the
massless Goldstone mode, which induces both supercurvature and
semi-classical perturbations. Because of the different normalization of
the supercurvature mode in supernatural inflation, $H_F \to 2R_0^{-1}$,
the supercurvature constraint~(\ref{super}) should read, in this case:
\begin{equation}\label{super2}
R_0H_T > 2\sqrt2\,\Big({D_l^{SC}/A_{SC}^2\over D_l^S/A_S^2}\Big)^{1/2} 
\approx 0.7\,,
\end{equation}
which is not trivially satisfied, even for degenerate minima. On the
other hand, the eigenvalue $\gamma=R_0^2m_T^2/2$ for the Goldstone
mode~\cite{quasi} induces a large semi-classical perturbation
(\ref{gamma}) unless 
\begin{equation}
R_0H_T < \sqrt{4A_S\over3}\,\Big({D_l^S/A_S^2\over 
D_l^C/A_C^2}\Big)^{1/4} \approx 0.02\,.
\end{equation}
It is clear that these two constraints cannot be accommodated
simultaneously. For values of $\Omega_0>0.9$, the bounds give
\begin{equation}
3\,(1-\Omega_0)\lesssim R_0H_T\lesssim0.008\,(1-\Omega_0)^{-1/2},
\end{equation}
which cannot be satisfied unless $\Omega_0\gtrsim0.98$, see
Fig.~\ref{fig8}. Therefore, the model is incompatible with
observations for $\Omega_0\lesssim0.98$. 

\subsection{Induced gravity open inflation}

This model was proposed in Ref.~\cite{Green} as a way of avoiding the
problems of classical motion outside the bubble. The inflaton field is 
trapped in the false vacuum due to its non-minimal coupling to
gravity, with coupling $\xi$. When the tunneling occurs it is left
free to slide down its symmetry breaking potential $V(\varphi) =
\lambda(\varphi^2-\nu^2)^2/8$. The expectation value of the inflaton
at the global minimum gives the Planck mass today: $M_{\rm Pl}^2 =
8\pi\xi\nu^2$. The model is parametrized by $\alpha =
8U_F/\lambda\nu^4$, which determines the value of the stable fixed
point in the false vacuum, $\varphi_{\rm st}^2 = \nu^2 (1+\alpha)$, 
as well as the difference in the rates of expansion in the false and true
vacua, $H_F^2=H_T^2\,(1+\alpha)/\alpha$, and the slow-roll parameters,
$\epsilon = 8\xi/(1+6\xi)\alpha^2$, $\eta =
8\xi\,(1-\alpha)/(1+6\xi)\alpha^2$, see Ref.~\cite{induced}. 

We will assume, for the $\sigma$ field, a tunneling potential of the 
type
\begin{equation}
U(\sigma) = {1\over4}\lambda'\sigma^2(\sigma-\sigma_0)^2 + \mu U_0
\Big[1-\Big({\sigma\over\sigma_0}\Big)^4\Big]\,,
\end{equation}
where $\sigma_0 = M\sqrt{2/\lambda'}$, $U_0=M^4/4\lambda'$ and $\mu\ll1$
for the thin-wall approximation to be valid. This potential gives a
tunneling parameter $b = (2\pi/3\lambda') M^3/H_T M_{\rm Pl}^2$, which
determines the relation between the mass of the $\sigma$ field in the
false vacuum and the rate of expansion there, $M =
H_F\,(1+1/\alpha)^{1/2}/\mu b$. Thanks to $\mu\ll1$, we can have $M\gg
H_F$ for values of $b\geq1$, which induces gravitational-wave
anisotropies that are well under control.

Furthermore, the induced gravity model seems to be truly open, since
the inflaton field $\varphi$ is static in the false vacuum and there
is thus no supercurvature mode associated with classical motion
outside the bubble, see Ref.~\cite{quasi}. Therefore the
constraint~(\ref{gamma}) does not apply, and there exists for this
model a range of parameters for which all contributions to the CMB
anisotropies are compatible with observations, see Ref.~\cite{JGB}.
However, the instanton may not take you to $\varphi_{\rm st}$ in the
true vacuum, but to a different value, closer to the minimum of the
potential, $\varphi=v$. In that case, the number of $e$-folds is
smaller than expected, and so is the value of $\Omega_0$. Such effects
should be taken into account for the determination of the model
parameters.

\subsection{Open hybrid inflation}

This model was proposed recently~\cite{GBL}, in an attempt to produce a
significantly tilted scalar spectrum in the context of open inflation,
in order to be in agreement with large scale structure~\cite{WS}. It is
based on the hybrid inflation scenario~\cite{hybrid,CLLSW}, which has
recently received some attention from the point of view of particle
physics~\cite{Guth,GBLW,LR,Pana,Dvali}, together with a tunneling field
that sets the initial conditions inside the bubble.

\begin{figure}[t]
\centering
\hspace*{-2mm}
\leavevmode\epsfysize=6cm \epsfxsize=8.5cm\epsfbox{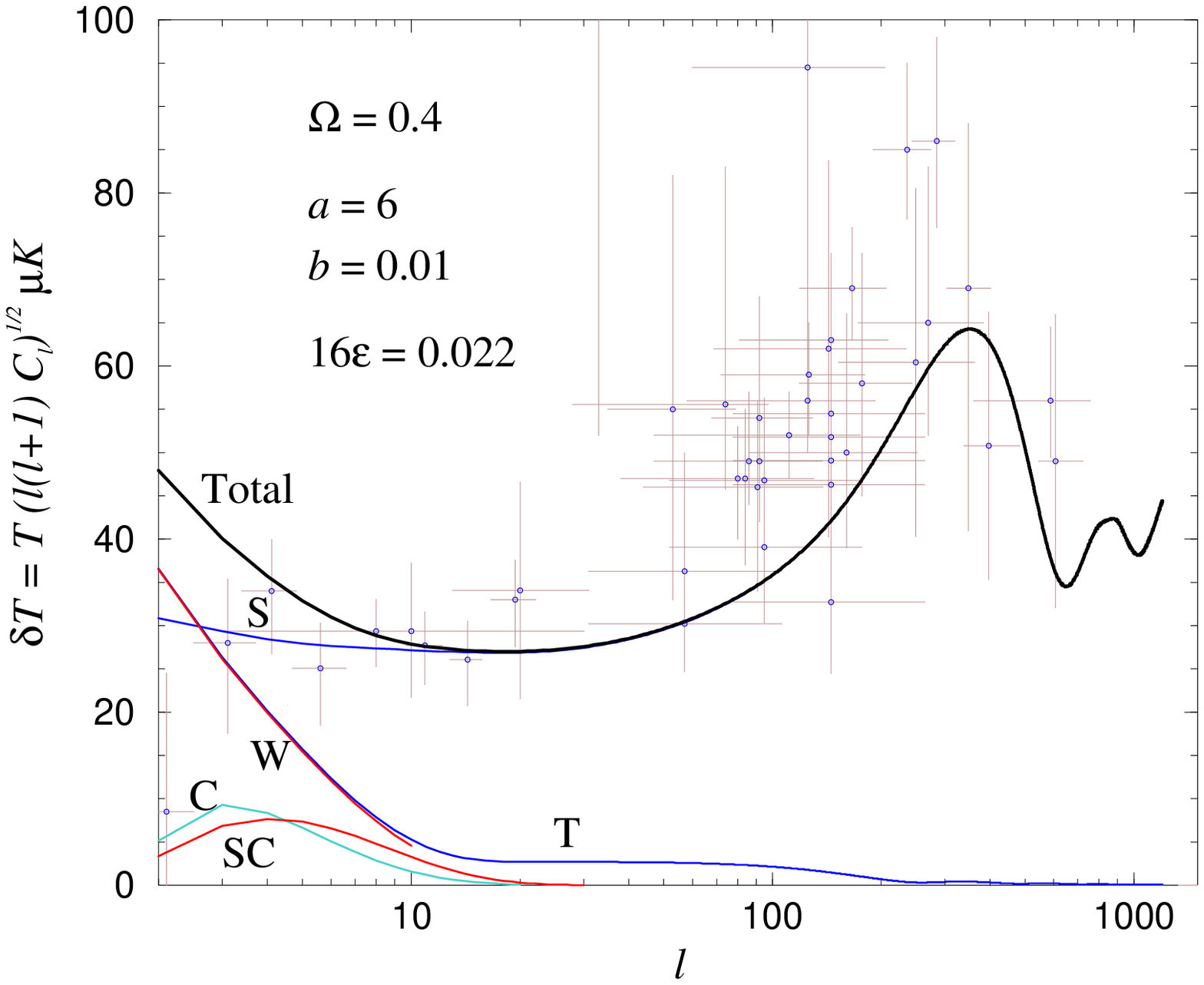}\\
\leavevmode\epsfysize=6cm \epsfxsize=8.5cm\epsfbox{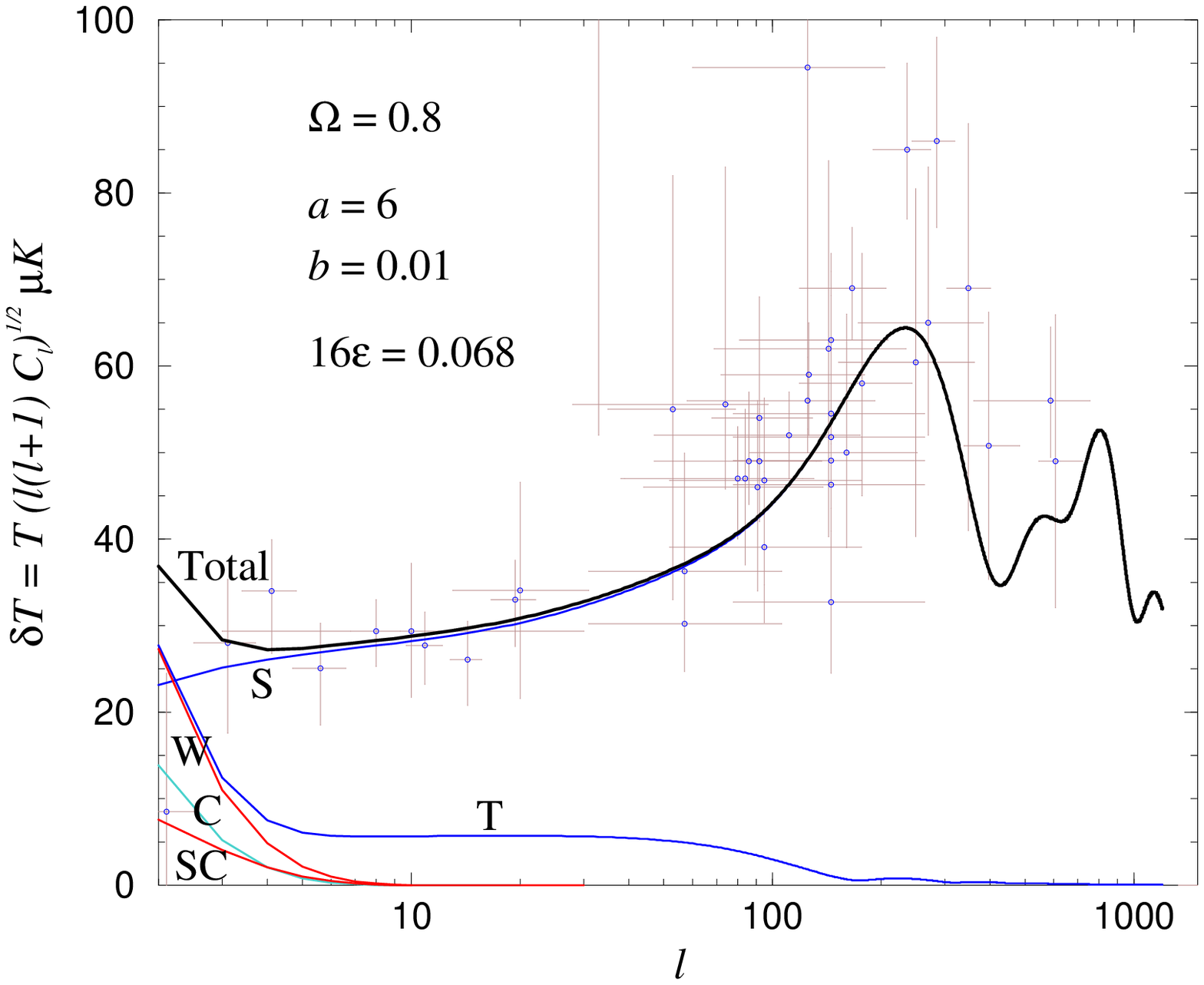}\\[3mm]
\caption[fig4]{\label{fig4} The complete angular power spectrum of
  temperature anisotropies for the open hybrid model with $\Omega_0 =
  0.4$ (higher plot) and $\Omega_0 = 0.8$ (lower plot), for $(a=6,
  b=0.01)$. The cosmological parameters are the same as in
  Fig.~\ref{fig5}. We show the individual contributions from the
  scalar, tensor, supercurvature, semi-classical and bubble-wall
  modes. Note that the bubble-wall mode is responsible for the large
  growth of the tensor contribution at low multipoles. Only the scalar
  modes remain beyond about $l=50$, where they grow towards the first
  acoustic peak.}
\end{figure}

In this model there are three fields: the tunneling field $\sigma$,
the inflaton field $\phi$ and the triggering field $\psi$. The
tunneling occurs as in the coupled model of Section \ref{twofields}
with potential
\begin{equation}\label{pot}
U(\sigma,\phi)= V_0+{\lambda\over4}\sigma^2(\sigma-\sigma_c)^2
+ {1\over2}g^2(\phi^2-\phi_c^2)\sigma^2 + U_0\,,
\end{equation}
where $\sigma_c = 2M/\sqrt\lambda$, \,$\phi_c = M/g$, \,$V_0\simeq
2.77M^4/\lambda$, to ensure that at the global minimum we have
vanishing cosmological constant, and $U_0$ is the vacuum energy
density associated with the triggering field. We satisfy $V_0\ll U_0$.
If the $\sigma$ field tunnels when $\phi=\phi_T=3\phi_c/4$, then
$\Delta U = U_F-U_T \simeq V_0/2 \simeq m_T^2\phi_c^2/4$.  After that,
the inflaton field will slow-roll down the effective potential
$U=U_0+m_T^2\phi^2/2\simeq U_0$ driving hybrid inflation, until the
coupling to $\psi$ triggers its end. The model is parametrized by
$\alpha = m_T^2/H^2_T$, see Refs.~\cite{GBL,JGB,Review}, in terms of
which the spectral tilts can be written as $n_S-1=2\alpha/3-6\epsilon$
and $n_T=-2\epsilon$. At tunneling we can write $V_0\simeq
m_T^2\phi_c^2/2=8m_T^2\phi_T^2/9$, so that the slow-roll parameter
$\epsilon = (\alpha/3) 9V_0/16U_T = 3\alpha ab/2$, where $4ab = \Delta
U/U_T\simeq V_0/2U_T$, see (\ref{ab}), and thus
\begin{equation}\label{ns}
n_S = 1 + {2\alpha\over3}\,\Big(1 - {27ab\over2}\Big).
\end{equation}

In order for open hybrid models to have a large tilt, we require {\em
both} a large value of $\alpha$ and a small value of $ab$. As we will
show, this will be impossible given the constraints
(\ref{bounds}--\ref{gamma}) from the CMB. For that purpose, we should
first compute the tunneling parameters $a=M/2 H_T$ and $b =
(4\pi\sqrt2/3\lambda) M^3/H_T M_{\rm Pl}^2 \simeq (V_0/4U_T)\,H_T/M =
2ab\,H_T/M$. Since both $H_T<M$ and $V_0\ll U_T$, we expect $a>1$ and
$b\ll1$, which will induce large tensor anisotropies at low
multipoles. This is a generic feature of open hybrid models. In
order to satisfy the CMB constraints we require
\begin{eqnarray}
H_F^2/H_T^2 &=& 1 + 4ab \lesssim 10 \,,\label{help1}\\
\epsilon &=& {3\alpha a b\over2} \lesssim 
{2b\,(0.6)\over[1+(a+b)^2]^{1/2}}  \,,\label{help2}\\
A_C &=& {3(1+4ab)\over16[1+(a+b)^2]^2} \lesssim 3\times 10^{-4}\,,\\
M &=& {2a\over(1+4ab)^{1/2}} H_F > H_F\,.
\end{eqnarray}
Note that $M_F^2\approx2M^2>2H_F^2$.  Since $a>1$ requires
$b<V_0/8U_T\ll1$, we can use the third constraint to get the bound
$a>5$, which then imposes (through the second constraint) that $\alpha
\lesssim 0.8/a(1+a^2)^{1/2}\simeq 1/a^2 < 1/30$. This means that the
scalar tilt (\ref{ns}) cannot be significantly larger than 1, as was
the aim of Ref.~\cite{GBL}.

We have plotted in Fig.~\ref{fig4} the complete angular power spectrum
of temperature anisotropies for the open hybrid model, in the case
$\Omega_0=0.4$ and $\Omega_0=0.8$, for $(a=6, b=0.01)$.
In order to prevent the tensor contribution from exceeding the cosmic
variance, we had to reduce the scalar spectral tilt to $n=1.002$,
which is essentially scale invariant and may not be sufficient to
allow consistency with the large-scale structure~\cite{WS}.
Furthermore, as we decrease in $\Omega_0$ it will be necessary for
scalar spectra to be closer and closer to scale invariance, in order
to reduce the tensor contribution. In any case, there exists for this
model a range of parameters for which all contributions to the CMB
anisotropies, see e.g. Fig.~\ref{fig4}, are compatible with
observations, even for low values of $\Omega_0$. Of course, as we
approach $\Omega_0\approx1$, it is much more likely to accommodate the
bounds.

\section{Probability distribution for $\Omega_0$}

As mentioned in Section IV, stronger constraints on two field models
arise if we take into account the probability distribution for
$\Omega_0$, which was considered in Ref.~\cite{gtv}. Inside a given
bubble, there will be observers which will measure all possible values
of $\Omega_0$, and the probability for a given value of $\Omega_0$ is
taken to be proportional to the number of collapsed objects of
galactic size that would form in all regions with that value of
$\Omega_0$. This probability is the product of three competing
factors. One is the ``tunneling'' factor, which basically corresponds
to Eq. (\ref{suppression}) and tends to suppress large values of
$\phi_0$, favouring low values of $\Omega_0$. The other is the
``anthropic'' factor, related to structure formation. The formation of
objects of galactic size is suppressed in a low density universe, and
so this factor favours large values of $\phi_0$.  Finally, there is
also a volume factor, taking into account that longer inflation leads
to more galaxies, although for model parameters where this factor is
dominant, the probability distribution is sharply peaked at
$\Omega_0=1$.

We should emphasize that we are assuming that the slow-roll field
outside the bubble does not affect the geometry of de Sitter space.
If, for instance, the Universe outside the bubble is in a process of
self-reproduction, the probability distribution for $\Omega_0$
inside the bubble may be affected. This issue requires further
investigation.

With the above assumptions, it was found that, in terms of the variable
$$
x\equiv \left({1-\Omega_0\over \Omega_0}\right),
$$
the logarithmic distribution $W=dP/d\ln x$
is peaked at the value
\begin{equation}
x_{\rm peak}\approx \kappa^{-1}\left({3\over 2} \mu -{5\over 4}\right)^{1/2},
\label{xpeak}
\end{equation}
where $\kappa \sim 0.1$ is a parameter related to structure formation,
and $\mu$ is given by:
\begin{equation}
\mu={\pi^2\over 6} R_0^4 V(\phi_0) \epsilon=
{1\over 16 A_S^2 [1+\Delta^2]^2}.
\label{mu}
\end{equation}
Near the peak value, the probability distribution $W$ is Gaussian, with
r.m.s. given by 
\begin{equation}
\Delta \ln x \sim (6\mu-5)^{-1/2}
\label{rmsx} 
\end{equation}
These expressions
are valid for $\mu \gtrsim 1$. For smaller values, the peak is at $x=0$,
meaning that most observers will see a flat universe.  On the other
hand, the value of $\mu$ should not be too large, say $\mu\lesssim 3$,
otherwise, from (\ref{xpeak}) and (\ref{rmsx}), we would have 
$\ln x_{\rm peak} \gtrsim 3$ and $\Delta \ln x \lesssim 0.3$, so
the observed value, $x\sim 1$, would be many standard deviations
away from the peak value.

Using $A_S \approx 5\times 10^{-5}$, and
Eqs.~(\ref{help1})--(\ref{help2}), we have $a\approx (4 A_S
\mu^{1/2})^{-1/2}, \ b\lesssim 4.5 A_S^{1/2} \mu^{1/4}$ and
\begin{equation}
\epsilon \lesssim 10^{-3} \mu^{1/2}.
\label{defcon}
\end{equation}
In the coupled model, $\epsilon \sim 10^{-2}$ and this constraint is
not satisfied for $\mu\lesssim 3$. One possibility would be to
increase the value of the parameter $\mu$, but then a value of
$\Omega_0\gtrsim 0.1$ would be extremely unlikely. Thus, the coupled
model does not seem to accomodate well an intermediate value of the
density parameter $0.1\lesssim \Omega_0 \lesssim 0.7$, and produce at
the same time sufficiently small CMB anisotropies. However, we must
recall that for values of $\Omega_0$ close to~1, the constraints from
supercurvature and bubble wall anisotropies are significantly
reduced. The constraints (\ref{help1}-\ref{help2}) can then be
substituted by (\ref{bsuperapp}-\ref{bwallapp}), and (\ref{defcon}) is
replaced by
$$
\epsilon \lesssim 4\times 10^{-6} (1-\Omega_0)^{-4} \mu^{1/2}.
$$
Thus, for $(1-\Omega_0)\sim 0.1$, the constraints are satisfied
even for $\epsilon\sim 10^{-2}$. 

\begin{figure}[t]
\centering
\hspace*{-2mm}
\leavevmode\epsfysize=6cm \epsfxsize=8.5cm\epsfbox{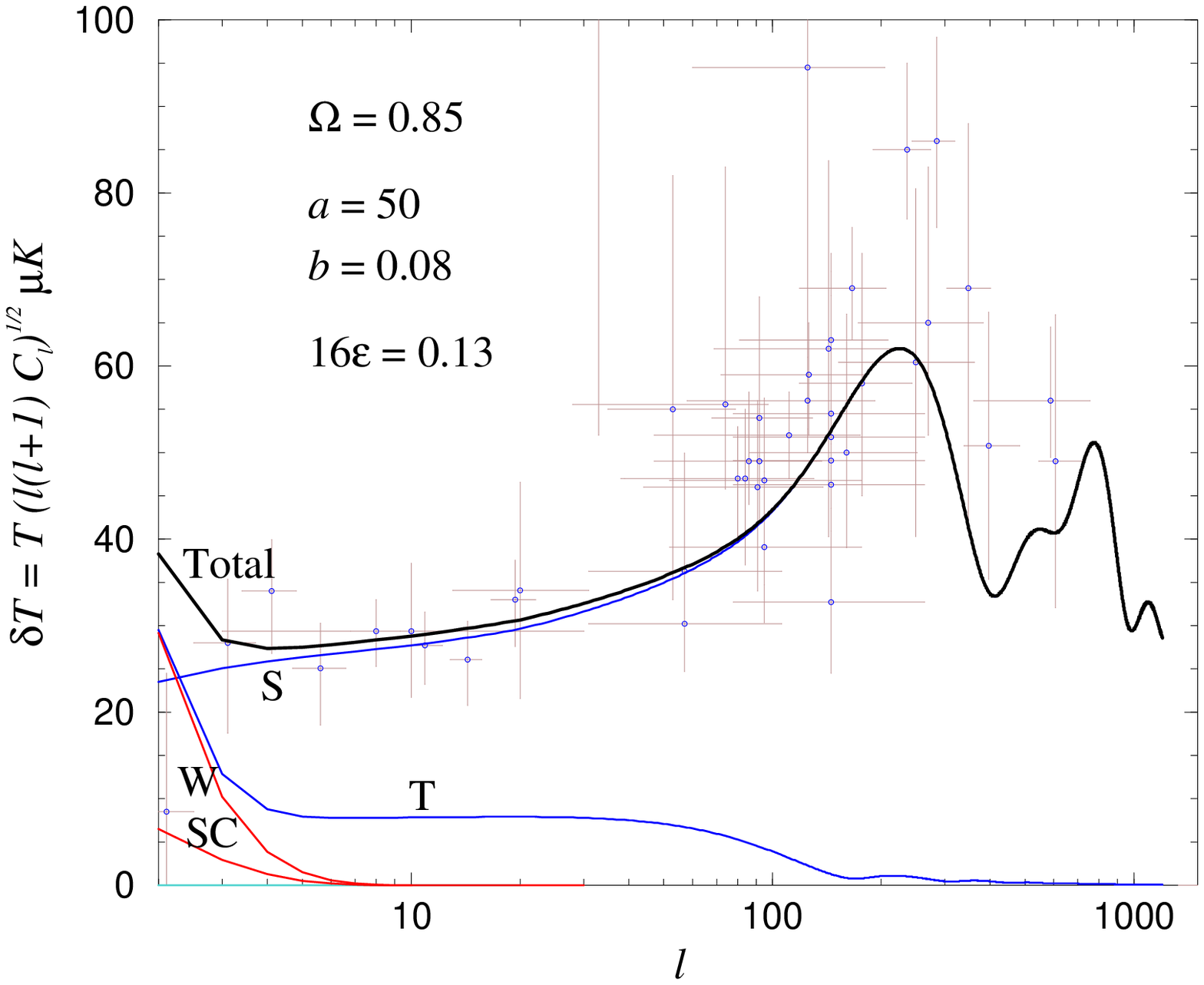}\\
\leavevmode\epsfysize=6cm \epsfxsize=8.5cm\epsfbox{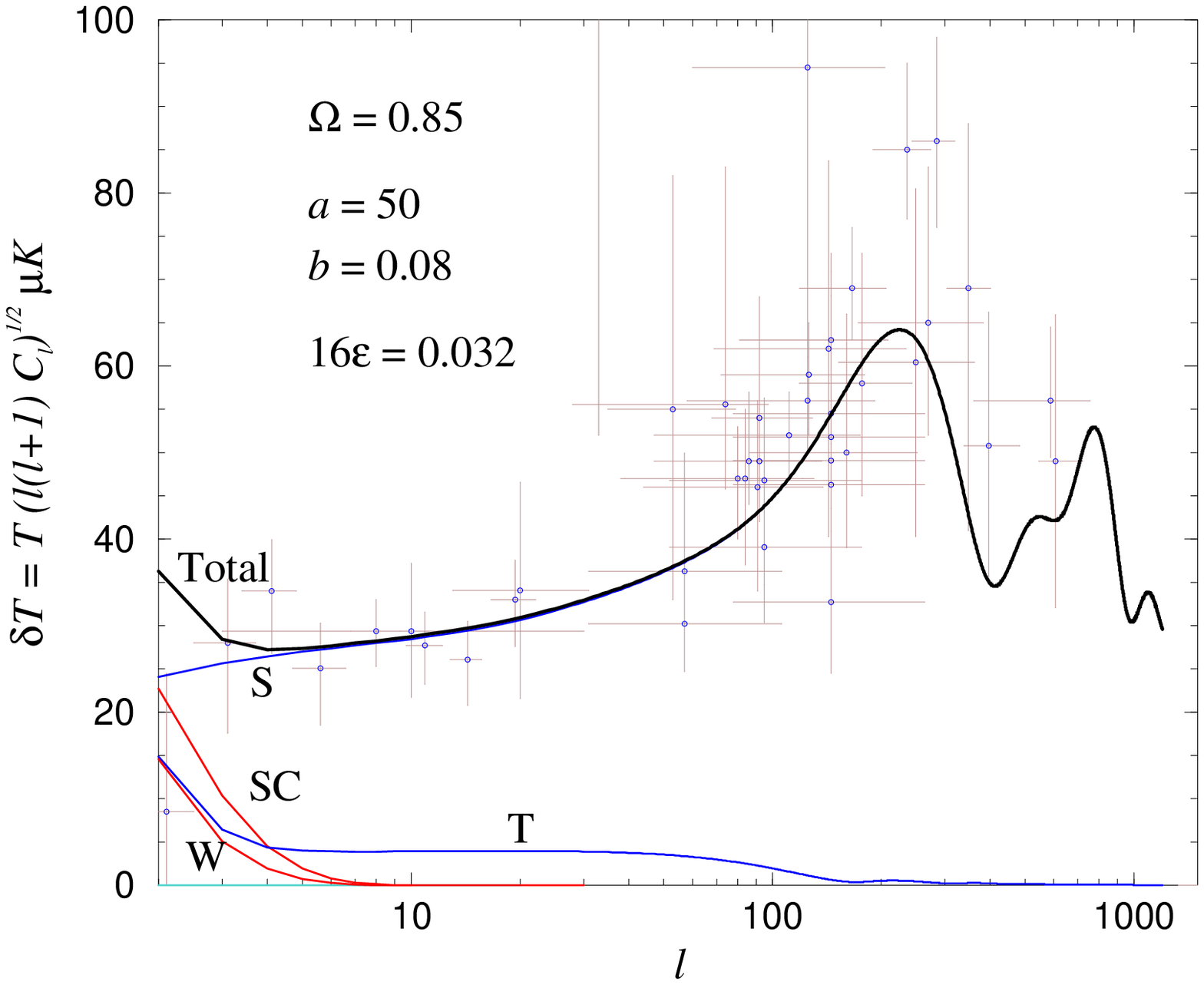}\\[3mm]
\caption[fig10]{\label{fig10} The complete angular power spectrum of
  temperature anisotropies for the coupled model (top panel) and for
  the open hybrid model (bottom panel) with $\Omega_0 = 0.85$, for
  $(a=50, b=0.08)$. The cosmological parameters are the same as in
  Fig.~\ref{fig5}.}
\end{figure}

For the sake of illustration, let us take $\mu=1$. Then the
probability distribution for $\Omega_0$ is peaked in the interesting
range: all values $0.1 \lesssim \Omega_0 \lesssim 0.9$ fall within one
standard deviation or so from the peak value and are not strongly
suppressed. This value of $\mu$ can be obtained by taking $a=50$ and,
for instance, $b=0.08$. Such values of $a$ and $b$ would lead to
unacceptably large supercurvature and wall fluctuation anisotropies if
we take, say $\Omega_0\lesssim 0.8$. However, for $\Omega_0\gtrsim
0.85$ we find that the anisotropies are below the observational bounds
(see Fig. \ref{fig10}). Therefore, the coupled model is in good shape
if the measured value of $\Omega_0$ turns out to be not too far from
1. This is very simple to understand: in that limit, all the effects
of the bubble wall are strongly suppressed.

This should be compared with the situation in Fig.~\ref{fig5}. There,
the CMB map is acceptable even for $\Omega_0=0.4$. However, with $a=10$,
we have $\mu\sim 10^3$. In this case the peak value is at $\ln x_{\rm
peak}\simeq6$ and the standard deviation is $\Delta\ln x \approx 0.01$.
This means that the ``measured'' value $\Omega_0 =0.4$ used for that
plot is formidably unlikely. For those values of the parameters, most
observers within the same bubble would measure much smaller values of
the density parameter.

Finally, for the open hybrid model~\cite{GBL}, the parameter $\epsilon$
can be made as small as desired, and the condition (\ref{defcon}) can be
easily satisfied. The reason is that in this model the range of values
of the inflaton field are well below Planck scale and the probability
distribution (\ref{prob}) easily covers those values within one
``standard deviation''.

\section{Conclusions}

Single-bubble open inflation is an ingenious way of reconciling an
infinite open Universe with the inflationary paradigm. In this
scenario, a symmetric bubble nucleates in de Sitter space and its
interior undergoes a second stage of slow-roll inflation to almost
flatness. At present there is a growing number of experiments studying
the CMB temperature anisotropies at fractions of a degree resolution
(corresponding to multipole ranges $l=20-500$), and they already put
some contraints on the spatial curvature of the universe.  However, in
the near future, observations of the CMB anisotropies with MAP and
Planck will determine whether we live in an open Universe or not with
better than 1\% accuracy. It is therefore crucial to know whether
inflation can be made compatible with such a Universe. Single-bubble
open inflation models provide a natural scenario for understanding the
large-scale homogeneity and isotropy, but most importantly, they
generically predict a nearly scale invariant spectrum of density and
gravitational wave perturbations. Future observations of CMB
anisotropies and large scale structure power spectra will determine
whether these models are still valid descriptions. For that purpose,
it is necessary to know the predicted spectrum with great accuracy.
In this paper we have explored the CMB anisotropy spectrum for various
models of single-bubble open inflation.  A host of features at low
multipoles due to bubble wall fluctuations, supercurvature modes and
quasiopenness place significant constraints on these models.

In particular, we find that the simplest uncoupled two-field model and
the ``supernatural'' model can only accomodate CMB observations
provided that $\Omega_0 \geq 0.98$.  Similarly, the simplest
single-field models of open inflation, based on a modification of new
inflationary potentials with the addition of a barrier near the
origin, induce too large tensor anisotropies in the CMB unless the
universe is sufficiently flat. Other single field models, where a
barrier is suitably appended to a generic slow-roll potential far from
the origin may not suffer from this problem \cite{newsingle}.

For the coupled two-field model, there is a range of parameters for
which all constraints from CMB anisotropies are satisfied even if
$\Omega_0$ is rather low (say $\Omega_0\approx 0.4$). On the other hand,
as argued in \cite{gtv} (see Section VI), stronger constraints arise if
we consider the probability distribution for the density parameter for a
given set of model parameters, and require that the measured value of
$\Omega_0$ is not too unlikely in the ensemble of all possible observers
inside the bubble. In that case, CMB constraints can still be
accomodated provided that $\Omega_0\geq 0.85$.

Finally, we have considered the open hybrid model, which was
introduced in \cite{GBL} with the motivation of generating a tilted
blue spectrum of density perturbations. In this model, all CMB
constraints can be accommodated even for low $\Omega_0$. Also, model
parameters can be chosen so that typical observers will measure the
density parameter in the range $0.1\lesssim \Omega_0 \lesssim 0.9$. In
this sense, the open hybrid model fares better than the coupled
two-field model. This is perhaps not too surprising, since this model
involves three fields and hence has more free parameters. Even so, it
turns out that for the parameter range where the CMB anisotropies are
compatible with observations at low multipoles, the tilt of the
scalar spectrum is negligible.

In conclusion, we find that existing models of open inflation are
strongly constrained by present CMB data. However, there is 
still for all of them a range of parameters where they would be 
compatible with observations.

Hawking and Turok \cite{HT} have recently proposed that it is possible
to create an open universe from nothing in a model without a false
vacuum.  The instanton describing this process is singular, and
therefore its validity has been subject to question
\cite{htall}. Nevertheless, it has also been pointed out that the
quantization of linearized perturbations in the singular background is
reasonably well posed \cite{jaume}.  Provided that one can make sense of
the instanton by appealing to an underlying theory where the singularity
is smoothed out, it seems that the details of that theory need not be
known in order to calculate the spectrum of cosmological
perturbations. This spectrum can be quite different from that of the
one-bubble universe case at large scales (see e.g. \cite{gmst2}, where
an analytically solvable model was considered), and it deserves further
investigation.

\section*{Acknowledgements}

It is a pleasure to thank Andrei Linde and Takahiro Tanaka for
useful discussions.  J.G.B. was supported by the Royal Society of
London, while J.G. and X.M.  acknowledge support from CICYT, under
contract AEN98-1093.

\appendix
\section{}

The open universe scalar harmonics for the sub\-curva\-ture modes can be
written as $Q_{qlm} = \Pi_{ql}(r)\,Y_{lm}(\theta,\phi)$, 
where~\cite{harrison}
\begin{equation}
\Pi_{ql}(r) = \sqrt{\frac{\Gamma(i q+l+1)\Gamma(-i q+l+1)}{\Gamma(i
    q)\Gamma(- i q)}}
{P_{iq-1/2}^{-l-1/2}(\cosh r)\over\sqrt{\sinh r}} \,.
\end{equation}
Here $Y_{lm}(\theta,\phi)$ are the usual spherical harmonics.

The open universe scalar harmonics for the supercurvature modes
($q^2=-\Lambda^2$) can be written as ${\cal Y}_{qlm}=\bar \Pi_{\Lambda,l}(r)\,
Y_{lm}(\theta,\phi)$, where~\cite{bubble}
\begin{equation}
\bar \Pi_{\Lambda,l}(r) =
\sqrt{{\Gamma(l+1+\Lambda)\Gamma(l+1-\Lambda)\over2}}
{P_{\Lambda-1/2}^{-l-1/2}(\cosh r)\over\sqrt{\sinh r}}\,.
\end{equation}
The various multipoles $l\geq \Lambda$ can be obtained from
\begin{eqnarray}
{P_{\Lambda-1/2}^{1/2}(\cosh r)\over\sqrt{\sinh r}}&=&\sqrt{2\over\pi}\,
{\cosh \Lambda r\over\sinh r}\,, \\
{P_{\Lambda-1/2}^{-1/2}(\cosh r)\over\sqrt{\sinh r}}&=&\sqrt{2\over\pi}\,
{\sinh \Lambda r\over \Lambda\sinh r}\,,
\end{eqnarray}
with the recurrence relation
\begin{eqnarray}
(\Lambda^2-l^2)\,P_{\Lambda-1/2}^{-l-1/2}(\cosh r)&=& \nonumber\\
P_{\Lambda-1/2}^{3/2-l}(\cosh r)-(2l-1)\coth r&&\hspace{-4mm}
P_{\Lambda-1/2}^{1/2-l}(\cosh r)\,.
\end{eqnarray}

To define the primordial scalar power spectrum we assume that
at the end of inflation the scalar metric perturbation takes the form
\begin{eqnarray}
{\cal R} &\to& \sum_{\Lambda} {\cal
  R}_{\Lambda} 
{\cal Y}_{\Lambda,lm} \\
&+& \sum_{\pm lm}\int_0^\infty dq \,{\cal R}_\pm(q)Q_{qlm} 
\end{eqnarray}
from where the explicit expressions for the amplitudes $A_S^2$,
$A_{SC}^2$ and $A_{W}^2$ can be read off. The continuum part of the
scalar power spectrum is defined as
\begin{equation}
\langle|{\cal R}(q)|^2\rangle = \sum_\pm|{\cal R}_\pm(q)|^2, 
\end{equation}
see \cite{YST} for details.

To describe the open universe gravitational waves we use the following
notation. The perturbed metric (if we only consider even
gravitational perturbations) can be written as 
\begin{equation}
ds^2 = a^2(\eta) \left[-d\eta^2 + 
(\gamma_{ij}^{(0)}+2h_{ij})dx^idx^j\right]\,.
\end{equation}
The perturbation $h_{ij}$ is then expanded as
\begin{equation}
h_{ij} = \sum_{\pm lm}\int_0^\infty dq\,h_{\pm qlm}(\eta) 
Q_{ij}^{qlm}(x^k)\,,
\end{equation} 
where the even harmonics $Q_{ij}^{qlm}(x^k)=G_{ij}^{ql}(r)
Y_{lm}(\theta,\phi)$ are transverse and
traceless\cite{harrison,TS,JGB}, and the radial component is given by
\begin{equation}
G_{rr}^{ql}(r) = \left[\frac{(l-1)l(l+1)(l+2)}{2
    q^2(1+q^2)}\right]^{1/2}\frac{\Pi_{ql}(r)}{\sinh^2 r}\,.
\end{equation}
At the end of inflation, the gravitational perturbation takes the form
\begin{equation}
h_{ij} \to \sum_{\pm lm}\int_0^\infty dq\,h_{\pm}(q) Q_{ij}^{qlm}(x^k)\,.
\end{equation}
The powers pectrum is then defined as
\begin{equation}
\langle|h(q)|^2\rangle = \sum_{\pm}|h_{\pm}(q)|^2\,,
\end{equation}
see \cite{TS,new} for details.

\section{}

As mentioned in Section IV, an observer located at a distance $r\gg 1$
from the center of an inflating island would measure an anisotropy due to
the gauge invariant perturbation (\ref{PHI}). This is
caused by a field perturbation 
$\delta\phi \approx (\gamma/2) \phi_0$, where $\phi_0$ is the value of
the scalar field at the beginning of inflation at the location where
the observer lives. We can compare this with the perturbation caused by 
the $l>0$ supercurvature modes, which is of order 
$\delta\phi \sim H_F/2\pi$. The
``semiclassical'' anisotropy would only dominate over the usual supercurvature 
anisotropies when 
\begin{equation}
\gamma \phi_0 \gtrsim H_F \label{pgh}.
\end{equation}
However, it turns out that such values of $\phi_0$ typically occur only
near the centers of inflating islands, as the following argument shows
\cite{gtv}. If $\phi_c$ is the value at the center, then
\begin{eqnarray}
\phi_0\approx \phi_c (1- {\gamma\over 6} r^2), \quad r\ll 1 \\
\phi_0\approx \phi_c (1- {\gamma\over 2} r),\quad  1\ll r \ll \gamma^{-1}
\label{sr}
\end{eqnarray}
Defining $\Delta \phi=\phi_c-\phi_0$, the probability for 
the observer to be at a distance $r$ away from the center can be obtained 
by expanding the exponent in (\ref{suppression})
$$
P(\phi_0+\Delta\phi) \propto \exp\Big[
{-\phi_0^2\over 2f^2}\Big(1+2{\Delta\phi\over\phi_0}\Big)\Big].
$$
Hence, the expected $\Delta\phi$ for an observer at $\phi_0$ is of order 
$\Delta\phi \lesssim f^2/\phi_0$.
Using (\ref{sr}), we see that for values of the field which
satisfy (\ref{pgh}), the expected distance to the center of the island is
of order
$$
r \sim {f^2\over \gamma\phi^2_0} \sim {H_F^2\over \gamma^2\phi^2_0} 
\lesssim 1.
$$
Therefore, one is led to the conclusion that field values for which
the semiclassical anisotropy would be large, satisfying (\ref{pgh}),
occur typically near the center of the islands, $r\ll 1$, where the 
anisotropy is actually not seen!

The same conclusion can be reached from first principles.  All the
necessary information is contained in the quantum state, a wave
functional depending on the amplitudes of the different field
multipoles. Expanding the field as $\phi(r,\theta,\varphi)=\sum c_{qlm}
Z_{qlm}(x^i)$, the square of the wave functional will give the
probability distribution $P[\phi]=\Pi_i P_i[c_i]$ for the coefficients
$c_i$, where $i=(q,l,m)$ is a collective index. $P$ factorizes
into independent $P_i$'s (which are just Gaussian distributions for each
$c_i$), because we are quantizing linearized perturbations which are
decoupled from each other.  The quantum state we are using is
homogeneous, and we can take any point on the hyperboloid as the origin
of coordinates.  Let us then take our observer to be at $r=0$, and let
us concentrate on the supercurvature sector $q^2=-1$. All modes with
$l>0$ vanish at the origin. Therefore, the value of the field at $r=0$
only depends on the coefficient $c_{q^2=-1,0,0}$ in our universe.  The
$l=0$ mode is spherically symmetric and hence does not contribute to
anisotropies. The anisotropies measured by this observer will only
depend on the amplitudes taken by the $c_i$ with $l>0$, whose r.m.s.  is
of order $H_F$. From this point of view, it is clear that typical
observers will effectively not see the semiclassical anisotropy
discussed in Section IV.

However, as discussed above, the ``weak'' assumption that our value 
of $\phi_0$ is not too special, in the sense that it will occur typically 
at large distances from the center of the island, implies that
$\gamma \phi_0 \lesssim H_F$. In other words, we must impose that the
anisotropy induced by the perturbation (\ref{PHI}) should always be
subdominant with respect to the usual supercurvature anisotropy.

\end{document}